\documentclass[twocolumn,showpacs,prb,footinbib,a4paper,superscriptaddress]{revtex4}

\usepackage{amssymb,amsmath}
\usepackage{graphicx}
\usepackage{dcolumn}
\usepackage{bm}

\newcommand{\expect}[1]{\langle #1 \rangle}
\newcommand{\evib}{{e\text{-vib}}} 
\newcommand{\elvib}{electron-vibron\ } 
\newcommand{\ph}{\text{ph}}
\newcommand{\GF}{\text{Green's function}\ }
\newcommand{\GFs}{\text{Green's functions}\ }

\begin{document}

\title{Non-equilibrium electronic structure of interacting
single-molecule nanojunctions: vertex corrections and polarization
effects for the electron-vibron coupling.}

\author{L.~K. Dash}
\affiliation{Department of Physics, University of York, York YO10 5DD, UK}
\affiliation{European Theoretical Spectroscopy Facility}
\author{H. Ness}
\affiliation{Department of Physics, University of York, York YO10 5DD, UK}
\affiliation{European Theoretical Spectroscopy Facility}
\author{R.~W. Godby}
\affiliation{Department of Physics, University of York, York YO10 5DD, UK}
\affiliation{European Theoretical Spectroscopy Facility}

 \date{\today}

\begin{abstract}
  We consider the interaction between electrons and molecular
  vibrations in the context of electronic transport in nanoscale
  devices.  We present a method based on non-equilibrium \GFs to
  calculate both equilibrium and non-equilibrium electronic properties
  of a single-molecule junction in the presence of \elvib interactions.
  We apply our method to a model system consisting of a single
  electronic level coupled to a single vibration mode in the molecule,
  which is in contact with two electron reservoirs.  Higher-order
  diagrams beyond the usual self-consistent Born approximation (SCBA)
  are included in the calculations.  In this paper we consider the
  effects of the double-exchange diagram and the diagram in which the
  vibron propagator is renormalized by one electron-hole bubble.  We
  study in detail the effects of the first- and second-order diagrams
  on the spectral functions for a large set of parameters and for
  different transport regimes (resonant and off-resonant cases), both
  at equilibrium and in the presence of a finite applied bias.  We
  also study the linear response (linear conductance) of the
  nanojunction for all the different regimes.  We find that it is
  indeed necessary to go beyond the SCBA in order to obtain correct
  results for a wide range of parameters.
\end{abstract}


\maketitle

\section{Introduction}\label{sec:introduction}

Single-molecule electronics has shown significant progress in the
recent years.  A variety of interesting effects have been observed in
the transport properties of single (or a few) conjugated organic
molecules including rectification, negative differential resistance,
and switching
\cite{Chen_J:1999,Rawlett:2002,Li:2003,Venkataraman:2006,Keane:2006,Venkataraman:2007}.
In these quasi-one dimensional systems, which present well delocalised
$\pi$-electrons, the electronic current flowing through the quite
flexible backbone of the molecule affects the ground state properties
of both electronic and mechanical degrees of freedom of the molecule.

The importance of inelastic effects in the transport properties has
been demonstrated in several ground-breaking experiments
\cite{Liu:2004,Beebe_JM:2007,Yu_L:2006,Kushmerick:2004,Yu:2004}; these
effects are related to the interaction between electron and mechanical
degrees of freedom of the molecule.

The interaction between an injected charge carrier (electron or hole)
and the mechanical degrees of freedom (phonon, vibron) in the
molecular junctions is important in order to understand energy
transfer, heating and dissipation in the nanojunction
\cite{Horsfield:2006}.  The \elvib interaction is also at the heart of
inelastic electron tunneling spectroscopy (IETS).  IETS is a
solid-state-based spectroscopy which gives information about the
vibration modes (vibrons) of the molecules in the nanojunction
\cite{Hipps:1993}.  It is now possible to measure such vibrational
spectra for single molecules by using scanning tunneling microscopy
(STM)~\cite{Liu:2004,Okabayashi:2008} to build IETS
maps~\cite{Gawronski:2009}, or by using other electromigrated
junctions or mechanically controlled break junctions
\cite{Beebe_JM:2007,Yu_L:2006,Kushmerick:2004,Yu:2004}.

At low applied bias (typically 100-400 meV) the IETS show features
(peaks, dips, or peak-dip-like lineshape) which have been attributed
to selective excitation of specific vibration modes of the molecule.
The position in energy (bias) of the features correspond approximately
to the frequency of the vibration, as given by other spectroscopic
data (IR, Raman) obtained on the same molecules in a different
environment.

There have been many theoretical investigations focusing on the
effects of electron-vibron coupling in
molecular and atomic scale wires \cite{Ness:1999, Ness:2001,
  Ness:2002, Flensberg:2003, Mii:2003, Montgomery:2003b, Troisi:2003,
  Chen:2004, Lorente:2000, Frederiksen:2004b, Galperin:2004,
  Galperin:2004b, Mitra:2004, Pecchia:2004, Pecchia:2004b,
  Chen_Z:2005, Paulsson:2005, Ryndyk:2005, Sergueev:2005, Viljas:2005,
  Yamamoto:2005, Cresti:2006, Kula:2006, Paulsson:2006, Ryndyk:2006,
  Troisi:2006b, Vega:2006, Caspary:2007, Frederiksen:2007,
  Galperin:2007, Ryndyk:2007, Schmidt:2007, Troisi:2007, Asai:2008,
  Benesch:2008, Paulsson:2008, Egger:2008, Monturet:2008,
  McEniry:2008, Ryndyk:2008, Schmidt:2008,
  Tsukada:2009,Loos:2009,Avriller:2009}.  Most of them have focused
on the interpretation of the features in IETS. However, most of these
studies have been performed by using the lowest-order expansion
possible for treating the effects of the \elvib interaction (i.e using
the so-called self-consistent Born approximation SCBA).  In the
language of many-body perturbation theory, it corresponds to a
self-consistent Hartree-Fock calculation for the \elvib coupling.

However, in analogy to what is obtained at the Hartree-Fock level for
interacting electrons, there are good reasons to believe that this
approximation is not enough to correctly describe the physics of the
\elvib interacting system, especially beyond the weak \elvib coupling
regime.  For example, the limits of SCBA have already been
investigated in Ref.[\onlinecite{Lee:2009}] but without introducing
remedies to go beyond SCBA.

In this paper, we examine this using a true non-equilibrium
Green's-function (NEGF) technique \cite{Mii:2003,Frederiksen:2004b,
  Galperin:2004b, Mitra:2004, Pecchia:2004b, Chen_Z:2005, Ryndyk:2005,
  Sergueev:2005, Viljas:2005, Yamamoto:2005, Cresti:2006, Vega:2006,
  Egger:2008} which allows us to study all the different transport
regimes in the presence of \elvib interaction.  Following the spirit
of many-body perturbation theory and Feynman diagrammatics, we go
beyond the commonly-used SCBA approximation by introducing
higher-order diagrams for the \elvib interaction.
 
We study the simplest possible model system which nonetheless contains
the relevant physics of the transport properties of the molecular
junction \cite{Lee:2009,Frederiksen:2007}.  Furthermore, because of
the uncertainty of the exact geometry of the single-molecule junction
in the experiments, there is a corresponding uncertainty about how to
model the coupling between the molecule and the electrodes and
correspondly for the potential drops at each molecule-electrode
contacts. Hence we take the quantitites characterizing the potential
drops at the contacts as phenomenological parameters
\cite{Galperin:2004,Datta:1997}.

We concentrate in this paper on the electronic properties of the
molecular junction in both equilibrium and non-equilibrium conditions
as well as on the linear-response properties of the junction (prior to
considering the full non-equilibrium transport properties in a
forthcoming paper).  Such properties are given by the density of
electronic states and represented by the spectral functions, which are
at the very heart of all physical properties of the system, such as
the charge density, the current density, the total energy, etc.

Spectral functions are most closely related to photoemission and
adsorption spectroscopies.  To our knowledge such experiments have not
yet been performed on single-molecule junctions, though photoemission
spectra have been measured on quasi one-dimensional supported atomic
scale metallic wires (see for example Ref.[\onlinecite{Nagao:2006}]
showing interesting results on one-dimensional collective electronic
excitations).

The paper is structured as follows.  We start with a description of
our model system in Section \ref{sec:hamiltonian} and a discussion of
the relevant underlying theory of non-equilibrium Green's functions in
section \ref{sec:non-equil-greens}.  Our calculated spectral functions
are presented in Section \ref{sec:res:spec_fnc}, where we consider
first the equilibrium case (Section \ref{sec:at-equilibrium}) and then
the non-equilibrium case (Section \ref{sec:NonEq-SpecFns}) at the
Hartree-Fock level. We discuss especially the effects of including or
not the Hartree diagram in the calculation.  We also compare NEGF
calculations with results obtained from inelastic scattering
techniques \cite{Ness:2006,Ness:2005,Ness:2001} for equivalent model
systems in Section \ref{sec:comp-with-inelastic-scattering-technique}.
We show that it is indeed necessary to go beyond SCBA to obtain
correct results for the relevant range of \elvib coupling.  We then
present the effects of the second-order diagrams in the spectral
functions in Section \ref{sec:vert-corr-spectr}. The second-order
diagrams correspond to two classes of process; the first is related to
vertex corrections of the SCBA calculation and the second to
polarisation effects (i.e. partial dressing of the vibron propagator
by the electron-hole bubble diagram).  In Section
\ref{sec:linear_reponse} we discuss the effects of different levels of
approximation for the \elvib coupling on the linear conductance of the
molecular junctions.  Throughout the paper, we will use the term
vibron to define a quantum of vibration of a mechanical degree of
freedom.

\section{Model}\label{sec:model}

\subsection{Hamiltonian}\label{sec:hamiltonian}

\begin{figure}
  \centering
   \includegraphics[width=\columnwidth]{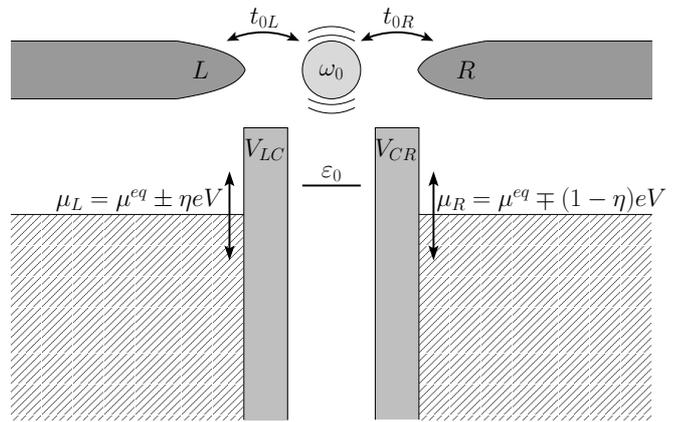} \\ 
   \caption{Schematic representation of the single site single mode
    (SSSM) model.  The single electronic level $\varepsilon_0$ is
    broadened by the coupling to the leads; the electronic transport
    is resonant if the Fermi levels $\mu_{L,R}$ are very close to this
    broadened level, and off-resonant (as shown) otherwise.  The
    system is shown in equilibrium, with $\mu_L = \mu_R$,
    non-equilibrium effects are studied by moving the Fermi levels.  }
  \label{fig:LeadsMolecule}
\end{figure}

Our model is based on a system with an interacting central region
connected to two non-interacting electrodes (see figure
\ref{fig:LeadsMolecule}).

The total Hamiltonian for the system is given by
\begin{equation}\label{eq:H_total}
  H_{\text{total}} = H_e + H_{\text{vib}} + H_{\evib},
\end{equation}
where $ H_e, H_{\text{vib}}$ and $H_{\evib}$ represent the
electronic, vibron, and electron-vibron coupling parts of the
Hamiltonian respectively. The electronic part of the Hamiltonian is
broken into sections describing the left (right) electrode $H_L$
($H_R$), the central interacting part $H_C^e$ and the potentials
coupling the central region to the left and right electrodes $V_{LC} +
V_{CR}$ respectively:
\begin{equation} \label{eq:H_electron}
  H_e = H_L + H_C^e + H_R + V_{LC} + V_{CR}.
\end{equation}

The Hamiltonians for the electrodes are given by
\begin{equation}
  \label{eq:H_electrodes}
  H_L + H_R = \sum_{\alpha = L,R} \varepsilon_\alpha c^\dag_\alpha c_\alpha,
\end{equation}
where $c^\dag_\alpha$ ($c_\alpha$) creates (annihilates) a
\emph{non-interacting} electron with energy $\varepsilon$ on electrode
$\alpha$.  The electronic Hamiltonian for the central region and the
coupling potentials are given by
\begin{gather}
  H_C^e = H_C^e (\{d^\dag_n\},\{d_n\}), \label{eq:H_central} \\
  V_{LC} + V_{CR} = \sum_{n,\alpha = L,R} V_{\alpha n} c^\dag_\alpha
  d_n + V^*_{\alpha n} d^\dag_n c_\alpha,  \label{eq:CouplingPot}
\end{gather}
where the interacting electrons in the central region are created
(annihilated) on electronic level $n$ by the operators $d^\dagger_n$
($d_n$). 

In principle, our Hamiltonian for the central region may contain
electron-electron interactions, and is built from a complete set of
single-electron creation and annihilation operators.  In our current
work we do not include any electron-electron interactions, and so the
electronic part of the total Hamiltonian for the central region $H_C$
becomes
\begin{equation}
  \label{eq:H_C_nonint}
  H_C^e= \sum_n \varepsilon_n d^\dag_n d_n.
\end{equation}

Meanwhile, the vibron part of the Hamiltonian is represented by
\begin{equation}
  \label{eq:H_vibron}
  H_{\text{vib}} = \sum_\lambda \hbar \omega_\lambda a^\dag_\lambda a_\lambda,
\end{equation}
where $ a^\dag_\lambda$ creates (annihilates) a vibron in vibron mode
$\lambda$ with frequency $\omega_\lambda$.  The electron-vibron
coupling term is taken to be linear in the vibron displacement and its
most general expression is then given by \cite{Ness:2001}
\begin{equation}
  \label{eq:H_electron_vibron}
  H_{\evib} = \sum_{\lambda,n,m} \gamma_{\lambda nm}
  (a^\dag_\lambda + a_\lambda) d^\dag_n d_m,
\end{equation}
where $\gamma_{\lambda nm}$ is the coupling constant for exciting 
the vibron mode $\lambda$ by electronic transition between the electronic 
levels $n$ and $m$.

We concentrate on the simplest version of the Hamiltonian of the
central part: the single-site single-mode SSSM model, in which one
considers just one electron level coupled to one vibration mode.
Despite the simplicity of this model, remarkably it not only contains
all the physics we require but also ensures that we isolate the
properties we are interested in without the complication of the added
electronic levels.  The reasons for this are as follows.  Firstly,
when the Fermi energy of the leads is pinned around the midgap, then
at low and intermediate biases one of the frontier orbitals (either
HOMO or LUMO) dominates the transport properties.  Secondly in
conjugated organic molecules (mostly used in single-molecule junction
experiments), it is known that the optically relevant vibration modes
are mostly coupled to either the HOMO or LUMO levels
\cite{Heeger:1988}.  We can therefore be confident that our model
contains the relevant physics.

The total Hamiltonian for the central region thus becomes
\begin{equation}
  \label{eq:H_c-sssm}
  H_C = \varepsilon_0 d^\dagger d + \hbar \omega_0 a^\dagger a +
  \gamma_0 (a^\dagger + a) d^\dagger d,
\end{equation}
where we now have just one electronic level $\varepsilon_n \rightarrow
\varepsilon_0$ and just one vibron mode $a_\lambda \rightarrow a$, coupled
via the \elvib coupling constant $\gamma_0$.  
The expression for the lead-central-region coupling (equation (\ref{eq:CouplingPot})) 
also simplifies to become
\begin{equation}
  \label{eq:SSSM-coupling-pot}
  V_{LC} + V_{CR} = \sum_{\alpha = L, R} t_{0\alpha} (c^\dagger_\alpha
  d + d^\dagger c_\alpha),
\end{equation}
where we have replaced the coupling potentials $V_{\alpha n}$ with
hopping integrals $t_{0\alpha}$.

\subsection{Non-equilibrium electron Green's functions and \elvib self-energies}
\label{sec:non-equil-greens}

Non-equilibrium \GFs (NEGF) within the Keldysh formalism
\cite{Keldysh:1965,Craig:1968,vanLeeuwen:2006,Rammer:2007} represent
an extremely useful tool for studying the non-equilibrium properties
of many-particle systems.  The \GFs are functions of two space-time
coordinates, and are obviously more complicated than the one-particle
density which is the main ingredient of density-functional-based
theories.  One of the great advantage of NEGF techniques is that one
can improve the calculations in a systematic way by taking into
account specific physical processes (represented by Feynman diagrams)
which is what we do in this paper for the \elvib
interaction.  The \GFs provide us directly with all expectation values
of one-body operators (such as the density and the current), and also
the total energy, the response functions, spectral functions, etc.

In Appendix \ref{app:higherorderdiagrams}, we provide more details
about NEGF and how to obtained the \elvib self-energies from a Feynman
diagrammatic expansion of the \elvib interaction. We now briefly
describe how we apply the NEGF formalism to the SSSM model.

The Green's functions are calculated via Dyson-like equations.  For
the retarded and advanced Green's functions $G^{r,a}$ these are
\begin{equation}
  \label{eq:Greens-retarded-advanced}
  G^{r,a} = g^{r,a}_C + g^{r,a}_C \Sigma^{r,a} G^{r,a},
\end{equation}
where $g^{r,a,}_C$ is the non-interacting Green's function for the
isolated central region.

For the greater $G^>$ and lesser $G^<$ Green's functions, we use a
quantum kinetic equation of the form
\begin{equation}
  \label{eq:Greens-greater-lesser}
  G^{>,<} = (1 + G^r \Sigma^r) g^{>,<}_C (1 + \Sigma^a G^a) + G^r
  \Sigma^{>,<} G^a.
\end{equation}
Here $\Sigma^x, (x = r, a, >, <)$ is a total self-energy consisting of
a sum of the self-energies from the constituent parts of the system:
\begin{equation}
  \label{eq:total-self-energy}
  \Sigma^x = \Sigma^x_L + \Sigma^x_R + \Sigma^x_{\text{int}}.
\end{equation}
$\Sigma^x_{L,R}$ are the self-energies arising from the
non-interacting leads $\alpha = L,R$ and as such are simple to
calculate:
\begin{gather}
  \label{eq:Lead-self-energies}
  \Sigma^r_\alpha = t_{0\alpha}^2 g^r_{0\alpha}(\omega), \\
  \Sigma^a_\alpha = (\Sigma^r_\alpha)^*, \\
  \Sigma^>_\alpha =  2i\ \Im m[\Sigma^r_\alpha(\omega)] \ (1 - f_\alpha(\omega)), \\
  \Sigma^<_\alpha = -2i\ \Im m[\Sigma^r_\alpha(\omega)] \ f_\alpha(\omega),
\end{gather}
where $f_\alpha$ is the Fermi-Dirac distribution for lead $\alpha$,
with Fermi level $\mu_\alpha=\mu^{\rm eq}+\eta_\alpha eV$ and 
temperature $T_\alpha$.
The fraction of potential drop at the left contact is
$\eta_L=\pm \eta_V$ and $\eta_R=\mp (1-\eta_V)$ at the right contact \cite{Datta:1997}, 
hence $\eta_L-\eta_R=eV$ is indeed the applied bias, and $\eta_V \in [0,1]$.

The component of the retarded Green's function for the isolated
(non-interacting) lead $\alpha$ corresponding to the sites (or energy
levels, depending on the representation used to the electrodes)
connected to the central region is given by $ g^r_{0\alpha}$.

In this paper, we have chosen a simple model, a
semi-infinite tight-binding chain with on-site energy
$\varepsilon_\alpha$ and nearest-neighbour hopping integral
$\beta_\alpha$.  This model gives a semi-elliptic density of states of
the terminal lead site connected to the central region, and so each
lead's Green's function becomes
\begin{equation}
  \label{eq:g0r_leads}
g^r_{0\alpha}(\omega) = \exp({\rm i}k_\alpha(\omega) ) /\beta_\alpha \,
\end{equation}
with
$\omega=\varepsilon_\alpha+2\beta_\alpha\cos k_\alpha(\omega)$.
We have chosen this model because it is one of the most simple, although
in principle and in practice there are no limitations for taking any
other more complicated or more realistic models for the lead, such as
Bethe lattices with $z$-coordination, or a nanotip supported by a
semi-infinite surface as shown in Ref~\onlinecite{Ness:2006})
since all their electronic properties are wrapped up in the lead
self-energies $\Sigma^x_{L,R}(\omega)$.

The self-energy for the interacting central region,
$\Sigma_\text{int}$ is somewhat more complicated.  It consists of the
sum of the self-energies due to interactions between the electrons
and to interactions between the electrons and the quantum vibration modes (vibrons).
In this paper we consider only the coupling between each electron
and a single vibron of the central
region (the molecule), hence the \elvib self-energy $\Sigma_\evib$.

\begin{figure}
  \centering
  (a) $\Sigma_\evib^{\text{Fock}} = $ \includegraphics[height=0.1\columnwidth]{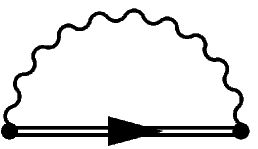}
  \ \ 
  (b) $\Sigma_\evib^{\text{Hartree}} = $\includegraphics[height=0.2\columnwidth]{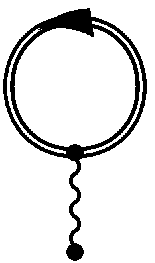}
  \caption{The (a) Fock and (b)  Hartree diagrams}
  \label{fig:FockHartree}
\end{figure}

\begin{figure}
  \centering
  (a) $\Sigma_\evib^{\text{DX}} = $ \includegraphics[height=0.18\columnwidth]{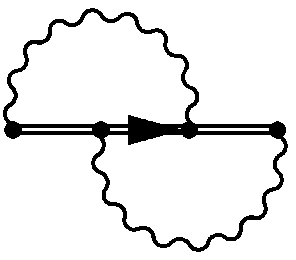}
  (b) $\Sigma_\evib^{\text{DPH}} = $ \includegraphics[height=0.15\columnwidth]{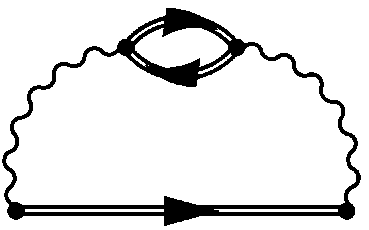}
  \caption{The (a) double exchange DX and (b) vibron propagator dressed by the e-h bubble diagrams (dressed phonon, DPH)}
  \label{fig:DX-DPH}
\end{figure} 

For the current work, it is necessary to calculate several types of
self energy.  Firstly we have the Fock-like self-energy
$\Sigma_{\evib}^{F,r/a/\gtrless}(\omega)$, which is a function of
energy, and is represented in diagrammatic form by
Figure~\ref{fig:FockHartree}(a). We also have the Hartree-like
self-energy $\Sigma_{\evib}^{H,r}$ which is independent of energy,
given by Figure \ref{fig:FockHartree}(b).  Calculations using only the
Hartree and Fock diagrams, and performed in a self-consistent way are
usually referred to as the self-consistent Born approximation (SCBA)
\cite{Mii:2003, Frederiksen:2004b, Galperin:2004, Mitra:2004,
  Pecchia:2004, Chen_Z:2005, Ryndyk:2005, Sergueev:2005, Viljas:2005,
  Yamamoto:2005, Cresti:2006, Vega:2006, Egger:2008}.

However, as explained in the introduction, we also want to go beyond
the SCBA, and thus we will also calculate two further self-energies
that include two-vibron processes.  The first of these is the
double-exchange self-energy $\Sigma^{\text{DX}}_\evib$ given by Figure
\ref{fig:DX-DPH}(a).  In the many-body language, it is part of the
vertex correction to the Fock diagram.  The second is given by Figure
\ref{fig:DX-DPH}(b) and corresponds to the dressed vibron (or GW-like)
self-energy $\Sigma^{\text{DPH}}_\evib$, which consists of the vibron
propagator renormalized by a single electron-hole bubble (the
polarization).  This is why we refer to the effects of
$\Sigma^{\text{DPH}}_\evib$ as polarization effects in the following.
The details of how we implement these self-energies are given in
appendix \ref{app:higherorderdiagrams}.

\subsection{Physical properties}
\label{sec:physical_prop}

Once we have calculated all the
different Green's functions, any of the physical properties of
the system, such as the electron density, the electronic current denstiy, 
the total energy, the current noise, the heat transfer, etc. can be calculated.  

For example, the electronic
current passing through the $\alpha$ contact is given by
\begin{equation}
  \label{eq:Current}
  I_\alpha(t)= \frac{2 {\rm i} e}{\hbar} \sum_n
  V_{\alpha n} \expect{c^\dag_\alpha(t) d_n(t)} - V^*_{\alpha n}
  \expect{d^\dag_n(t) c_\alpha(t)}, 
\end{equation}
i.e the first term describes the transfer of an electron from the
interacting region to electrode $\alpha$, while the second transfers
an electron from electrode $\alpha$ to the central region.

We can then express this in terms of Green's functions and derive an
expression for the expectation value of the current\cite{MeirWingreen:1992}:
 \begin{equation}\label{eq:Expectation_value_current}
    \begin{split}
      I_\alpha &= \frac{2e}{\hbar} \int \frac{{\rm d}\omega}{2\pi}\ 
    {\rm Tr}\{ 
    \Sigma_\alpha^<(\omega)\ G^>(\omega)-\Sigma_\alpha^>(\omega)\
    G^<(\omega)\} \\
    &= \frac{{\rm i} 2e}{\hbar} 
     \int \frac{{\rm d}\omega}{2\pi}
    {\rm Tr}   \left \{  f_\alpha(\omega) \Gamma_\alpha(\omega)
       [G^r(\omega)-G^a(\omega)] \right.  \\
      & \qquad\quad  \left.   + \ \Gamma_\alpha(\omega)\ G^<(\omega) \right \} \ .
  \end{split}
\end{equation}

All physical properties may be expressed in terms of the spectral
function $A(\omega)$ which is at the heart of this paper.  The
spectral function is related to the imaginary part of the retarded or
advanced electron Green's functions, as
\begin{equation}
\label{eq:spectral_function}
A(\omega)=- \Im m [G^r(\omega)] = +\Im m [G^a(\omega)] \ .
\end{equation}
For non-interacting systems, it is simply proportional to the density of
electronic states $n(\omega)= A(\omega)/\pi$.
For interacting systems, it gives information about the excitations (electron
or hole) of the system.

Furthermore, when the system is at equilibrium ($f_L=f_R=f^{\rm eq}$), there
are some relationships between the lesser (greater) and the advanced and retarded 
Green's functions:
\begin{equation}
\label{eq:Glesser_equi}
\begin{split}
G^{<,{\rm eq}}(\omega) = - f^{\rm eq}(\omega)\left( G^{r,{\rm eq}}(\omega) - G^{a,{\rm eq}}(\omega) \right) \\
= 2{\rm i} f^{\rm eq}(\omega) A(\omega),
\end{split}
\end{equation}
and
\begin{equation}
\label{eq:Ggreater_equi}
\begin{split}
G^{>,{\rm eq}}(\omega) = - (f^{\rm eq}(\omega)-1)\left( G^{r,{\rm eq}}(\omega) - G^{a,{\rm eq}}(\omega) \right) \\
= 2{\rm i}  (f^{\rm eq}(\omega)-1)A(\omega).
\end{split}
\end{equation}
These relationships are at the centre of the fluctuation-dissipation theorem
for equilibrium, which also be recast as a relationship betweeen the greater and
lesser Green's functions:
\begin{equation}
\label{eq:Glessgreat_KMS}
G^{>,{\rm eq}}(\omega)\ =\ - {\rm e}^{(\omega-\mu_0)/kT}\ G^{<,{\rm eq}}(\omega)
\end{equation}
for statistical averages at finite temperature in the grand canonical ensemble.
This equation is related to the Kubo-Martin-Schwinger boundary conditions 
\cite{Kadanoff:1962,vanLeeuwen:2006}.

For non-equilibrium conditions, there is no unique Fermi level at finite bias (or no 
unique temperature if $T_L\ne T_R$) in the whole system, and the relationships given 
by Eqs. (\ref{eq:Glesser_equi}-\ref{eq:Glessgreat_KMS}) no longer hold. 
This is an important feature of the non-equilibrium formalism for which conventional 
equilibrium statistics need to be reformulated.

\subsection{Computational aspects}\label{sec:some-comp-aspects}

The calculations start by constructing the non-interacting \GFs of the
entire system $G_0^{x}$ with $x$ being any three of the possible \GFs
$x=\{r,a,<,>,t,\tilde t\}$.  The other \GFs are obtained by using the
relationships between them as shown in
Appendix~\ref{app:1st_order_diagrams}.  For example, the retarded \GF
is given by $G^r_0(\omega)=[\ g^r_0(\omega)^{-1}-\Sigma^r_{\rm
  L}(\omega)-\Sigma^r_{\rm R}(\omega)]^{-1}$.

One then calculates three ``Keldysh components'' for the self-energies
corresponding to any of the diagrams $\bigstar=$ \{F,DX or DPH\}:
$\Sigma_\evib^{\bigstar,<}$, $\Sigma_\evib^{\bigstar,>}$ and
$\Sigma_\evib^{\bigstar,t}$.  The Hartree diagram has only one
component $\Sigma_\evib^{H,r}$ as shown above.

The advanced and retarded self-energies $\Sigma_\evib^{\bigstar,\{r,a\}}$ are then
obtained by simple algebra using the relationships between the
different self-energies as explained in Appendix~\ref{app:1st_order_diagrams}.

The new \GFs are then calculated by using the Dyson equations for $G^{r,a}$ 
\begin{equation}
  \label{eq:totalGFDyson}
  G^{r,a}(\omega) = [ \omega -\varepsilon_0 - \Sigma_\text{total}^{r,a}(\omega) ]^{-1},
\end{equation}
and the quantum kinetic equations for $G^{<,>}$
\begin{equation}
  \label{eq:totalGFKeldysh}
  G^\gtrless(\omega) = G^r(\omega)
  \Sigma_\text{total}^\gtrless(\omega) G^a(\omega) ,
\end{equation}
where the total self-energies are given by
\begin{equation}
  \label{eq:totalselfenergy}
  \Sigma_\text{total}^{x}(\omega) = \Sigma_L^{x}(\omega) +
  \Sigma_R^{x}(\omega) + \sum_{{\rm any } \bigstar}\Sigma_\evib^{\bigstar,x}(\omega) .
\end{equation}
The new self-energies are then recalculated and the process re-iterates until 
full self-consistency is achieved.

Actually at each iteration of the calculations, we use a simple mixing scheme
of the self-energies obtained at the present iteration and at the previous
iteration. This mixing scheme permits us to achieve full self-consistency in a
maybe slightly longer but more stable iterative process.

Note finally that by using Eq.~\ref{eq:totalGFKeldysh}, instead of the more
general formulation given by Eq.~\ref{eq:Greens-greater-lesser},
we assume that, after switching on the interactions, there are no bound
states in the system (i.e. there are no interaction-induced electron states
located outside the spectral supports of the left and right leads)
\cite{Thygesen:2007,Stefanucci:2007} which is indeed the case.

\section{Results: Spectral functions}\label{sec:res:spec_fnc}

In this section, we present results for the spectral
functions $A(\omega)$ for the
different transport regimes and for different applied
biases in the low vibron-temperature regime.

We divide the calculations into four types. Firstly, we calculate the
spectral functions at equilibrium for two transport regimes.  The
first of these is where $\varepsilon_0 \gg \mu^{\rm eq}$ or
$\varepsilon_0 \ll \mu^{\rm eq}$, known as the off-resonant regime as in
order to create a current between the left and right leads one puts an
electron in the empty (for electron transport, $\varepsilon_0 \gg
\mu^{\rm eq}$) or full (for hole transport, $\varepsilon_0 \ll
\mu^{\rm eq}$) electronic level $\varepsilon_0$.

The second transport regime is when 
$\varepsilon_0\pm\text{linewidth}\sim\mu^{\rm eq}$, known as the resonant transport regime.
The linewidth is the width of the peak in the spectral function
which arises from the electronic coupling of the central region
to the left and right leads.  In this case the electronic level
$\varepsilon_0$ is, at equilibrium, half-filled by an electron (and thus also
half-filled with a hole).

For each of these transport regimes, we calculate the spectral
function both at equilibrium (applied bias $V=0$) and non-equilibrium at
finite bias ($V>0$).

For each of these four groups (resonant/off-resonant transport
regimes, at/out of equilibrium), a large amount of different NEGF
calculations have been performed for different values of the relevant
parameters and within different levels of approximation (Hartree Fock,
Hartree Fock+second-order, partially or fully self-consistent
calculations).  In the rest of this section, we present only a limited
and selected number of results which we found the most relevant for
each case, and we analyse and compare in detail the effects of the
different diagrams on the spectral functions of the system at and out
of equilibrium.

We also, in section
\ref{sec:comp-with-inelastic-scattering-technique}, compare perturbation expansion 
based calculations (NEGF) to a reference calculation which is exact in
term of \elvib coupling but which however is only valid
for a specific transport regime.

\subsection{At equilibrium}
\label{sec:at-equilibrium}

We first consider the spectral functions at equilibrium, with no
applied bias.  Figure \ref{fig:specdens_equi_offres} shows $A(\omega)$
for the off-resonant regime, and Figure \ref{fig:specdens_equi_res}
shows the resonant transport regime.

In the equilibrium many-body language, the features in the spectral functions
obtained at positive energies ($\omega \ge 0$, above the Fermi
level) correspond to electron excitations, while the features
at negative energies ($\omega \le 0$, below the Fermi
level) correspond to hole excitations.

\subsubsection{Off-resonant transport regime}
\label{sec:SpectralFunctionsOffRes}

\begin{figure}
  \includegraphics[width=8cm]{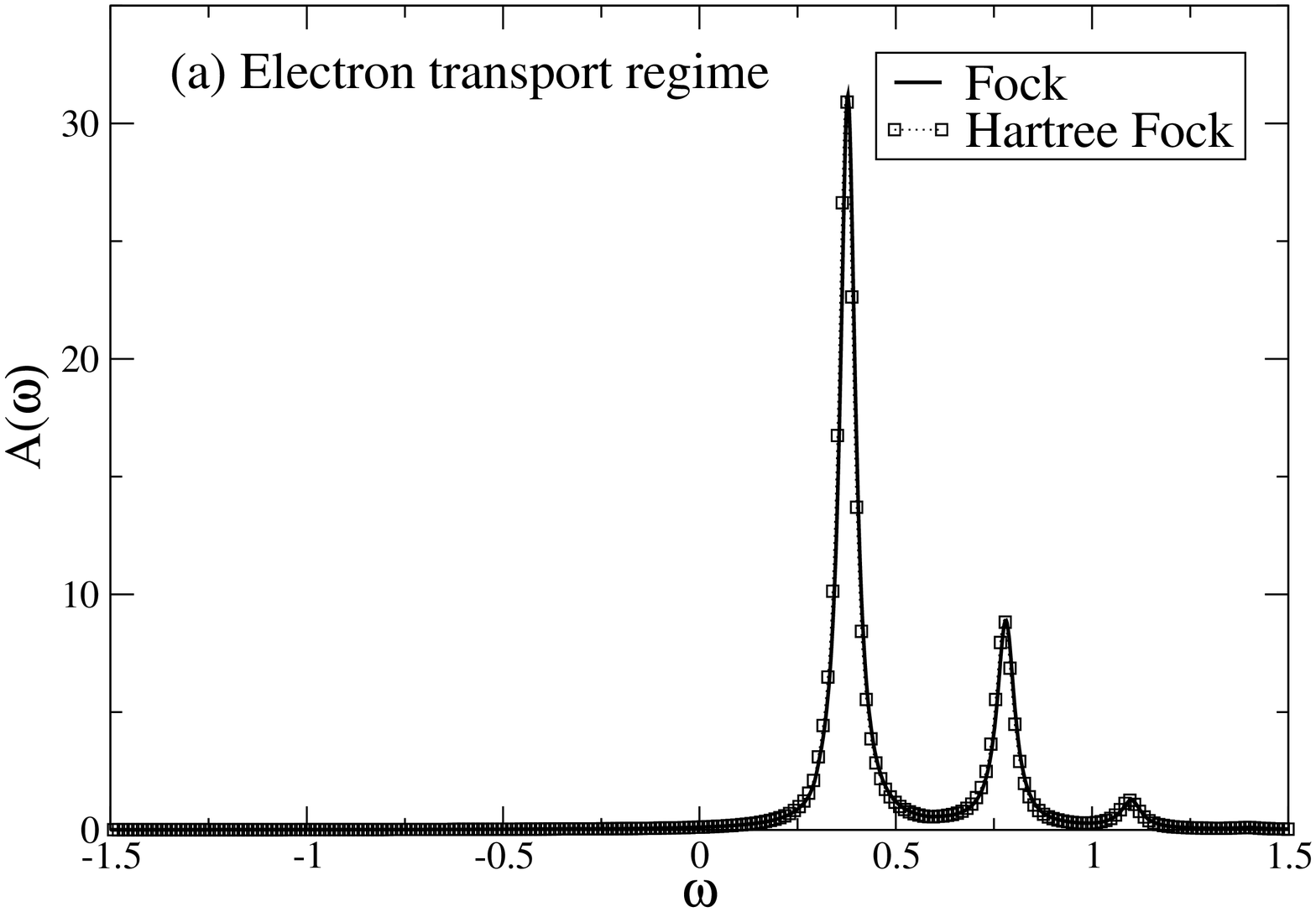}
   \\ 
    \vspace{2\parskip}
    \includegraphics[width=8cm]{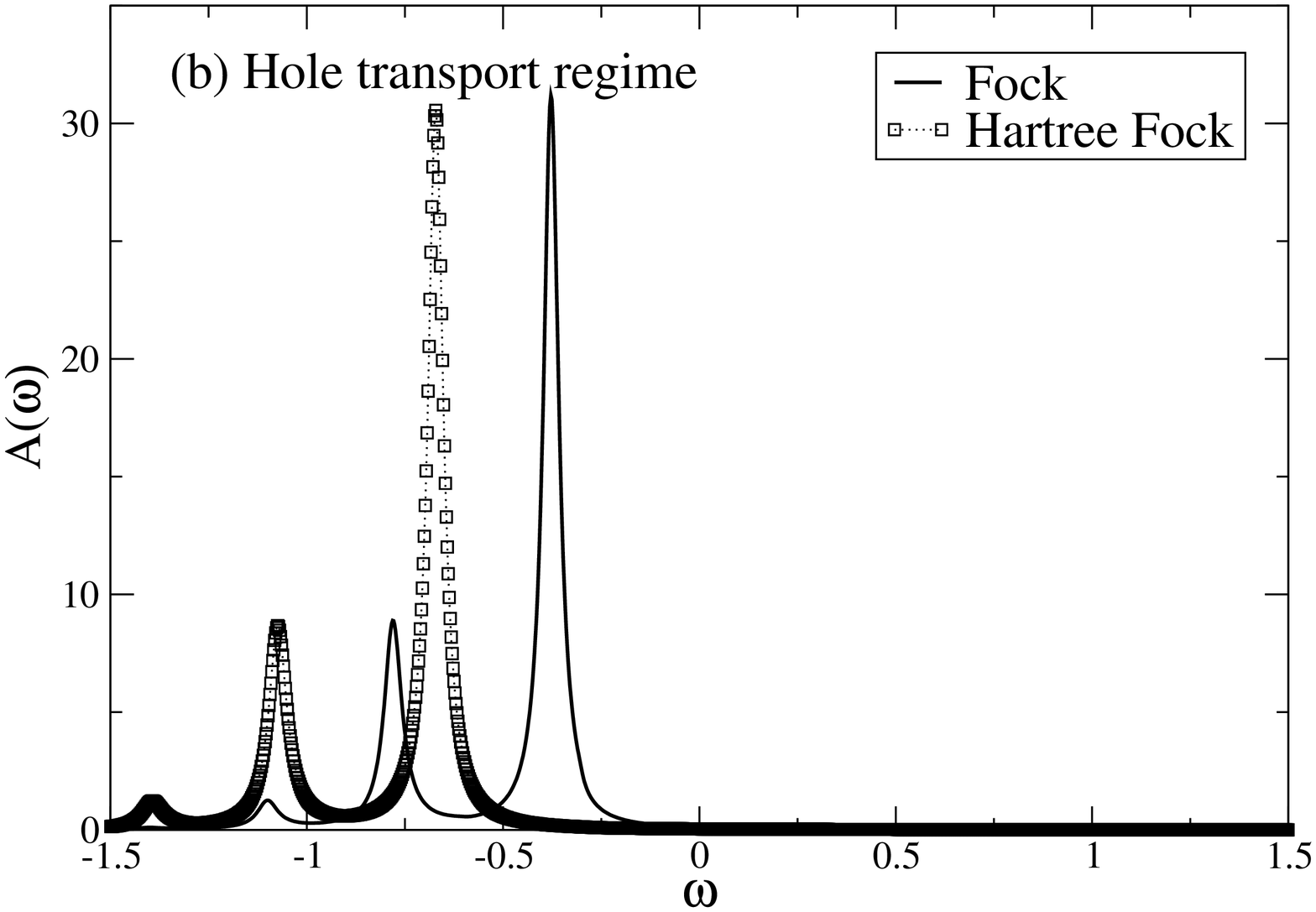}
    \caption{ Equilibrium (zero bias) spectral functions
      $A(\omega)$ for the off-resonant
      (a) electron and (b) hole transport regime.  Calculations were
      performed with the Fock-like \elvib diagram (solid line) and with
      both the Hartree and Fock-like diagrams (symbols and dotted
      line). For the electron transport regime the inclusion of the
      Hartree self-energy has no effect, but for the hole transport
      regime it shifts the entire spectral function to lower energies.
      The parameters are $\varepsilon_0=+0.5 (-0.5)$ for electron
      (hole) transport $\gamma_0=0.21, \omega_0=0.3, t_{0L,R}=0.15,
      \eta=0.005$
      and $\mu_L=\mu_R=\mu^{\rm eq}=0$.  }
    \label{fig:specdens_equi_offres}
\end{figure}

For the off-resonant electron transport regime
(Figure~\ref{fig:specdens_equi_offres}(a)) all the features in the
spectral function are above the Fermi level $\mu^{\rm eq}$ and hence
correspond to electron excitations.  The main peak corresponds to
adding an electron in the single available level.  This peak is
broadened by the coupling to the leads, and its position in energy
$\tilde{\varepsilon}_0$ is renormalised by the electron-vibron
interaction, i.e. $\tilde{\varepsilon}_0$ is close to the static the
polaron shift $\tilde{\varepsilon}_0 \approx \varepsilon_0 -
\gamma_0^2/\omega_0$.

The lesser peaks in the electron-transport spectral function are
vibron side-band peaks arising from resonance with $n=1,2,3,...$
excitations in the vibration mode.  These peaks correspond to
vibration excitation (vibron emission) only. At zero vibron
temperature, these are the only available mechanisms for vibrational
excitations.  We note, and discuss further in section
\ref{sec:comp-with-inelastic-scattering-technique}, that these
side-band peaks should occur at integer multiples of $\omega_0$ away
from the main peak, but that for both our Fock-only and Hartree-Fock
SCBA calculations the peak-peak separation is slightly wider than this.

In this regime, the Hartree self-energy is negligible, because
most of the spectral weight is above the Fermi level
and $\Sigma^H_\evib \propto \int^{\mu^{\rm eq}} {\rm d}\omega A(\omega) \sim 0$.
This implies that the polaron shift is mainly due to the Fock-like
self-energy.

For the off-resonant hole-transport regime (
Figure~\ref{fig:specdens_equi_offres}(b)), all the features in the
spectral function occur at $\omega < 0$ and therefore correspond to
hole excitations.  The vibron side band peaks are at lower frequencies
than the main peak because they correspond to the emission of vibrons
by holes rather than electrons.

When we include just the Fock-like self-energy, the hole spectral
function is symmetric (with respect to the equilibrium Fermi level
$\mu^{\rm eq}$) with the electron spectral function, as can be seen
clearly in figure~\ref{fig:specdens_equi_res}.  Adding the
Hartree self-energy, however, breaks this electron-hole symmetry.  As
most of the spectral weight is below $\mu^{\text{eq}}$, the expression
for $\Sigma^H_\evib$ given in equation Eq.(\ref{eq:Hartreeselfenergy})
reduces to a constant $2 \gamma_0^2/\omega_0$ (i.e. twice the polaron
shift) as $\int \frac{d\omega}{2\pi} \rm{i} G^<(\omega) \sim 1$.  As a
result of this the whole spectral function is shifted to the left by
this amount.

\subsubsection{Resonant transport regime}
\label{sec:SpectralFunctionsResonant}

\begin{figure}
  \includegraphics[width=8cm]{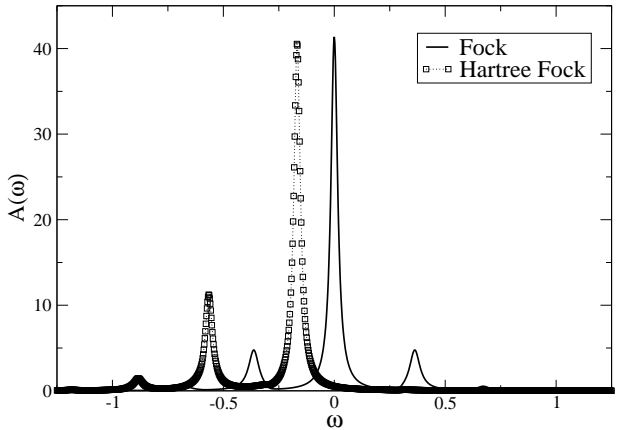}
  \caption{Equilibrium spectral function $A(\omega)$ for the 
    resonant transport regime. Calculations were performed with the
    Fock-like (solid line) and with both the Hartree and Fock-like 
    diagrams (dashed line). The parameters are identical to those 
    used in Figure \ref{fig:specdens_equi_offres}, except for the value 
    of the electron level which is $\varepsilon_0=0$ for the resonant 
    regime.}
    \label{fig:specdens_equi_res}
\end{figure}

We now turn to the resonant transport regime, with the spectral
function shown in Figure \ref{fig:specdens_equi_res}.  Here our
electronic level $\varepsilon_0$ is broadened by the coupling to the
leads and is partially filled with electrons.  We can see that
calculations performed with only the Fock-like self-energy preserve
the electron-hole symmetry. The spectral function presents peaks
located both at positive and negative energies, which correspond to
the emission of vibrons by electrons or holes respectively.  As in the
previous section, the inclusion of the Hartree self-energy (equation
~\eqref{eq:Hartreeselfenergy}) breaks down the electron-hole symmetry
and the features are shifted to lower energies with a corresponding
modification of the spectral weights for each peak.

We note here that the inclusion of the Hartree self-energy thus
modifies so drastically the spectral function that it will also
strongly affect the $I$-$V$ characteristics of the junction in
comparison to calculations performed with only the Fock self-energy
\endnote{L. K. Dash, H. Ness, R. W. Godby, unpublished.}.

\subsection{Non-equilibrium spectral functions}
\label{sec:NonEq-SpecFns}

\subsubsection{Non-equilibrium off-resonant transport}
\label{NonEq-SpecFns-OffRes}

In this section, we present results for the spectral
functions $A(\omega)$ for the
off-resonant transport regimes, for different applied
biases, both with and without the Hartree contribution.
We first present calculations for an asymmetric potential drop with
$\eta_V = 1$,
i.e. $\mu_L=V$ and $\mu_R=0$ ($\mu^{\rm eq} = 0$). 

\begin{figure}
  \includegraphics[width=8cm]{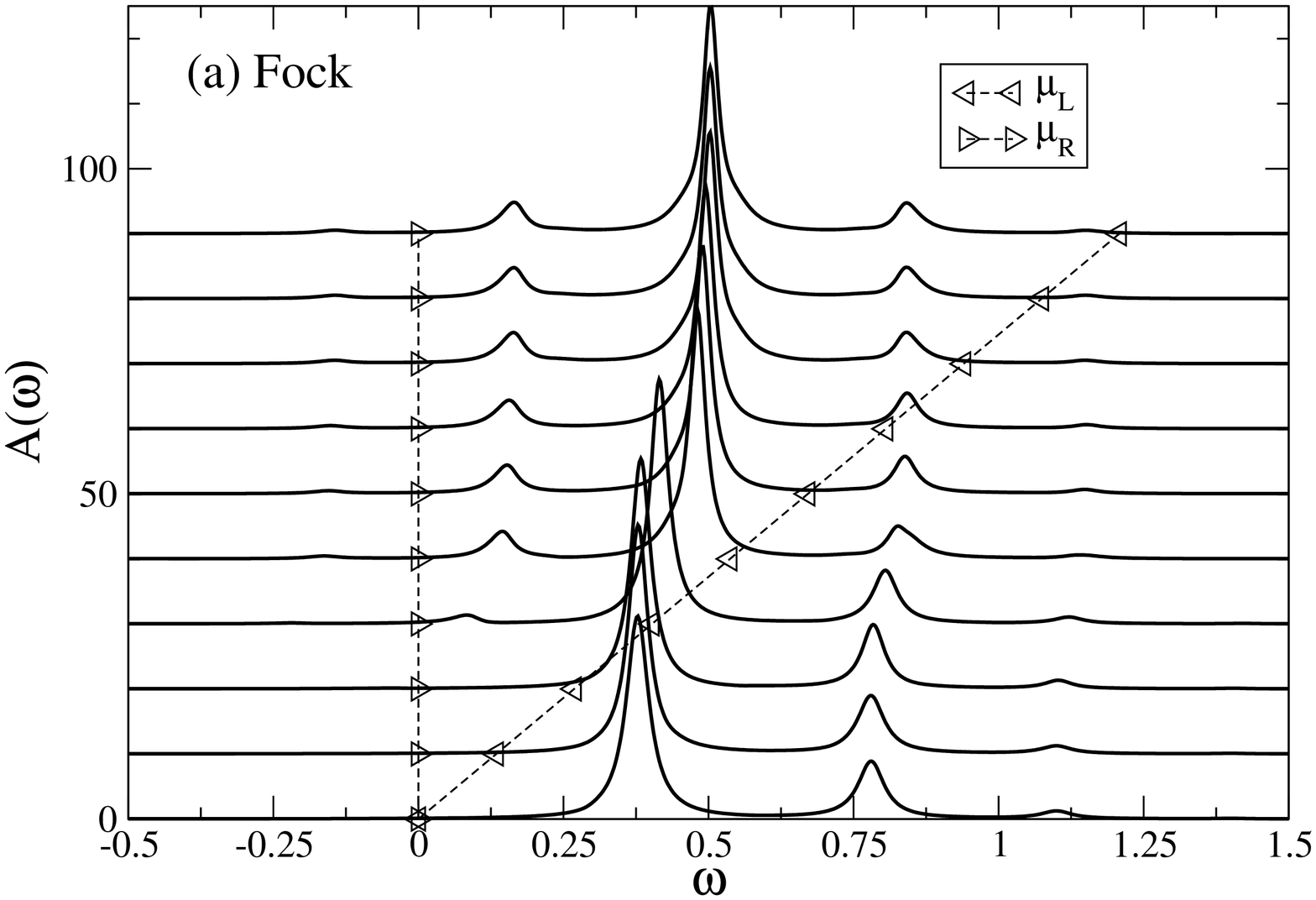}
  \\ 
  \vspace{2\parskip}
  \includegraphics[width=8cm]{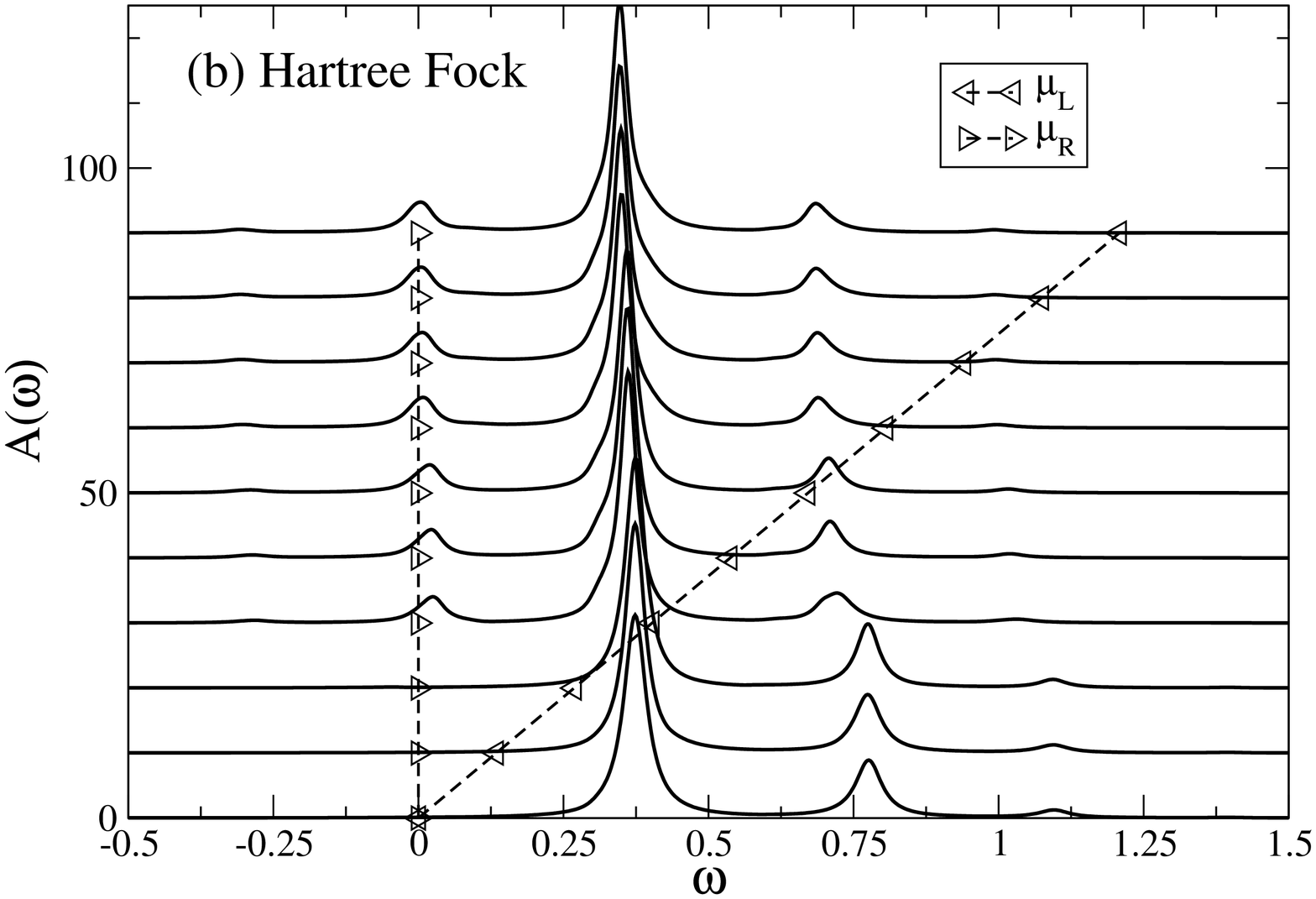}
  \caption[]{Non-equilibrium spectral function for the off-resonant
    transport regime with applied bias for (a) the Fock diagram only
    and (b) both Fock and Hartree diagrams.  The curves are offset
    vertically for clarity.  The applied bias is given by the chemical
    potentials of the left (left-pointing arrows) and right
    (right-pointing arrows) leads respectively. The Hartree potential
    has a strong effect on the peak positions.  The other parameters are
    $\varepsilon_0=+0.5, \gamma_0=0.21, \omega_0=0.3, t_{0L,R}=0.15,
    \eta=0.005, \eta_V=1$.}
  \label{fig:specdens_NE_offres_F}
\end{figure}

Figure \ref{fig:specdens_NE_offres_F}(a) shows the spectral function
calculated with just the Fock component at different applied biases.
We then increase the bias $V$ by increasing the chemical potential of
the left contact while keeping that of the right contact at zero.  For
low values of $V$ ($V \leq \omega_0$) there is little change in the
spectral function.  However, once the value of $V$ exceeds that of
$\omega_0$, the spectral function becomes increasingly modified,
especially when there is significant spectral weight inside the bias
window $\mu_R < \omega < \mu_L$. In particular, the lineshapes of both
the main peak and the vibron side-band peaks become deformed, with a
noticeable asymmetry of the main peak for biases where the main peak,
but not the right-hand vibron side-band peak, lies within the bias
window.

There is a saturation regime once the vibron side band peaks, and thus
nearly all of the spectral weight, is within the bias window (curves
for $\mu_R \gtrsim 1$).  Here the main peak becomes pinned around $\omega
= 0.5$ (effectively midway between $\mu_L$ and $\mu_R$), and a
symmetric lineshape is restored.  We postulate that this is owing to
the bias being large enough to achieve simultaneous electron and hole
transport.

Figure \ref{fig:specdens_NE_offres_F}(b) shows the spectral function
with both the Hartree and Fock diagrams included.  As we have already
noted, at zero bias there is no change from the Fock-only spectral
function.  For small biases ($V \lesssim \omega_0$) there is little
difference, but for larger biases the effect of the Fock diagram
becomes increasingly evident.  This is because once the bias exceeds
the vibron frequency $\omega_0$, the non-equilibrium electron density
becomes strongly perturbed.  The main peak is more stable in position
than for the Fock-only spectral function, although it shifts slightly
towards lower energies with increasing bias before stabilising.  The
righthand vibron side-band peak becomes strongly deformed, and moves
to lower, rather than higher, energies as the bias is increased.  The
left-hand vibron side-band peak appears at a lower bias than for the
Fock-only spectral function, at a frequency just above zero, then
tends towards zero as the bias increases.
As for the Fock-only case
however, once both vibron side-band peaks are within the bias window,
saturation is reached and the peak positions stabilize.  For both the
Fock-only and Hartree-Fock spectral functions, the separation between
both side-band peaks and the main peak is $\sim 0.34, > \omega_0$ for 
the parameters we used.

\begin{figure}
  \includegraphics[width=8cm]{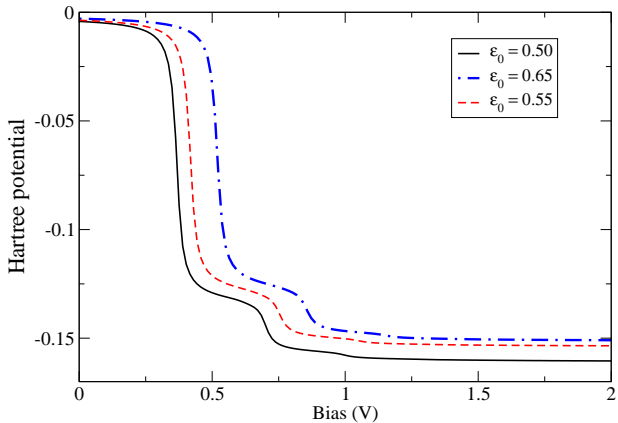}
  \caption{(color online) Hartree potential versus applied bias.  The Hartree
    potential is almost constant in the low bias regime, and hence it
    could be neglected in the calculations, because it just
    corresponds to a global energy reference shift. However for larger
    biases, it is much more influential as the non-equilibrium
    electron density varies a lot with the applied bias.  The
    parameters are the same as in figure
    \ref{fig:specdens_NE_offres_F}.}
  \label{fig:Hartreepot_vs_V_offres}
\end{figure}

Figure~\ref{fig:Hartreepot_vs_V_offres} shows the Hartree potential
(i.e.~the real part of $\Sigma_\evib^{H,r}$) plotted against applied
bias for different values of the electronic level $\varepsilon_0$.
The solid line corresponds to the value of $\varepsilon_0$ used to
calculate the spectral functions of the lower panel in
Figure \ref{fig:specdens_NE_offres_F}.  The Hartree potential is
small and almost constant in the low bias region.  Hence, in the
quasi-equilibrium regime, such a potential could be neglected in the
calculations, because it simply corresponds to an energy reference
shift.  However when the bias window $[ \mu_L , \mu_R ] $ starts to
encompass features in the spectral functions (either the main peak or
vibron side-band peaks), the corresponding non-equilibrium electron
density starts to vary substantially with the applied bias. Thus the
Hartree potential shows a strong dependence on the value of the bias
(as also shown in Ref.~[\onlinecite{Lee:2009}]),
until the saturation regime is reached and the Hartree potential is once
again  constant for very large biases.  The asymptotic saturation
value of the Hartree potential is dependent on the value of the
electronic level $\varepsilon_0$ as one would expect.   
The first drop in the value of the
Hartree potential happens at $\tilde{\varepsilon}_0$, with the
subsequent steps, which become progressively smaller and  broader, at
$\tilde{\varepsilon}_0 + \sim 0.34$ 
and then $\tilde{\varepsilon}_0 + n\omega_0$.

This behaviour indicates that one cannot in principle neglect the
Hartree diagram contribution in the calculations, unless one is
interested in calculating only the properties of the system for a bias
range for which the electron density is (almost) constant
\endnote{It is interesting to note that all the steplike features in
  the Hartree potential versus applied bias occur at the same biases
  as the steplike features in the current (not shown in this paper).
  In other words, peaks in the dynamical conductance $G(V)=dI/dV$ and
  in the derivative of the Hartree potential versus $V$ are obtained
  for the same bias.  Though, these two quantities contain the same
  spectral information, there is no simple relationship between them.
  Even in the off-resonant regime, the current
  Eq.~(\ref{eq:Expectation_value_current}) is not only given by the
  lesser \GF from which the Hartree potential is derived.}.

It is worth mentioning that it is not straightforward to relate the
modification of the peak positions in the two panels in
figure~\ref{fig:specdens_NE_offres_F} to the Hartree potential alone.
In self-consistent calculations, a highly non-linear system needs to
be solved, since the Hartree potential is obtained from one element of
the \GFs which are themselves dependent on the values of the Hartree
potential.

\subsubsection{Non-equilibrium resonant transport}
\label{NonEq-SpecFns-Resonant}

We now consider what happens when we apply a bias in the resonant
transport regime.  We apply a symmetric potential drop (i.e. the
potential of the left electrode is raised by an amount $\eta e V$
while that of the right electrode drops by the same amount). This
allows us to keep the electron-hole symmetry and see under which
circumstances the electron-hole symmetry is broken.  

\begin{figure}
  \includegraphics[width=8cm]{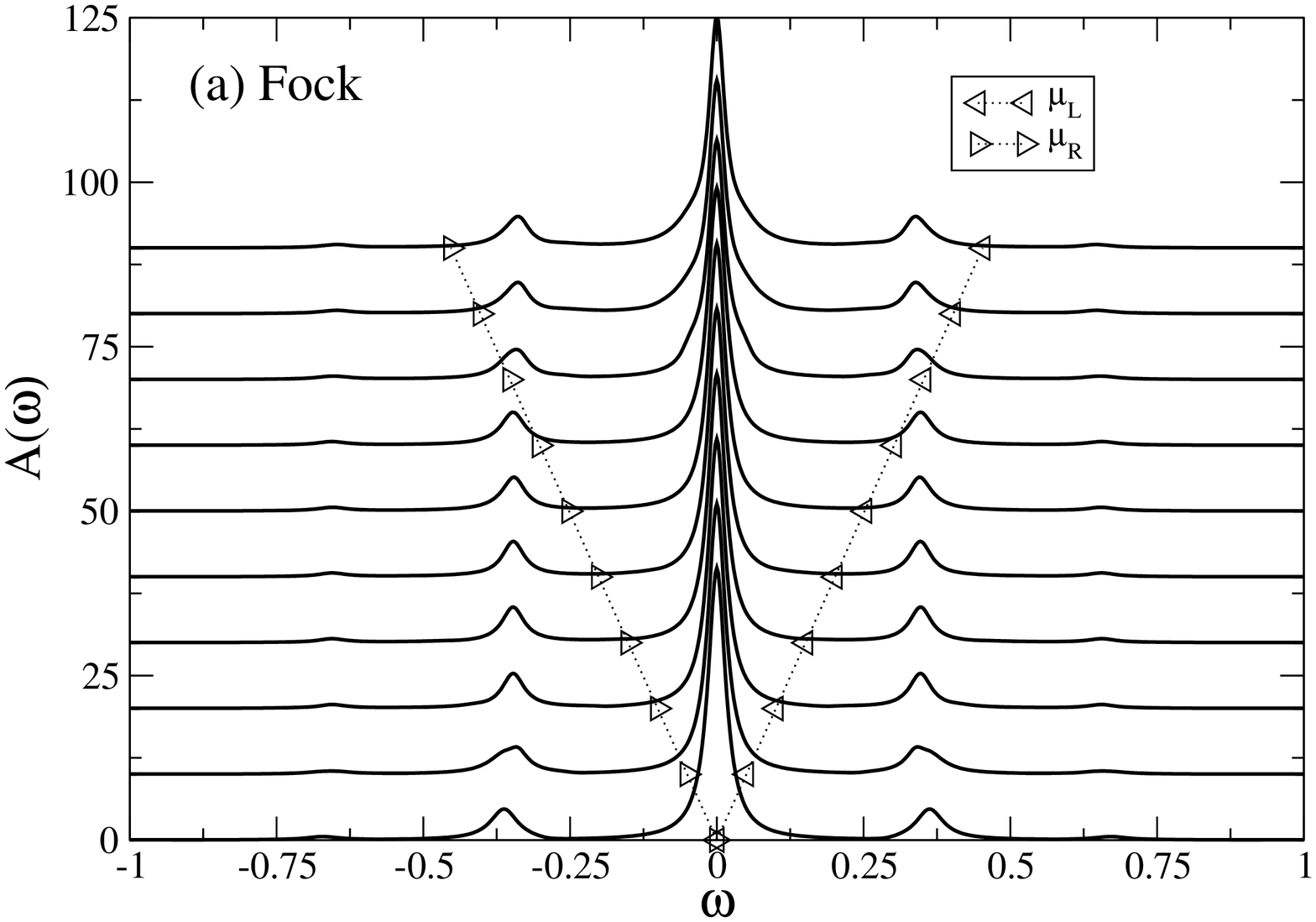} 
  \\
  \includegraphics[width=8cm]{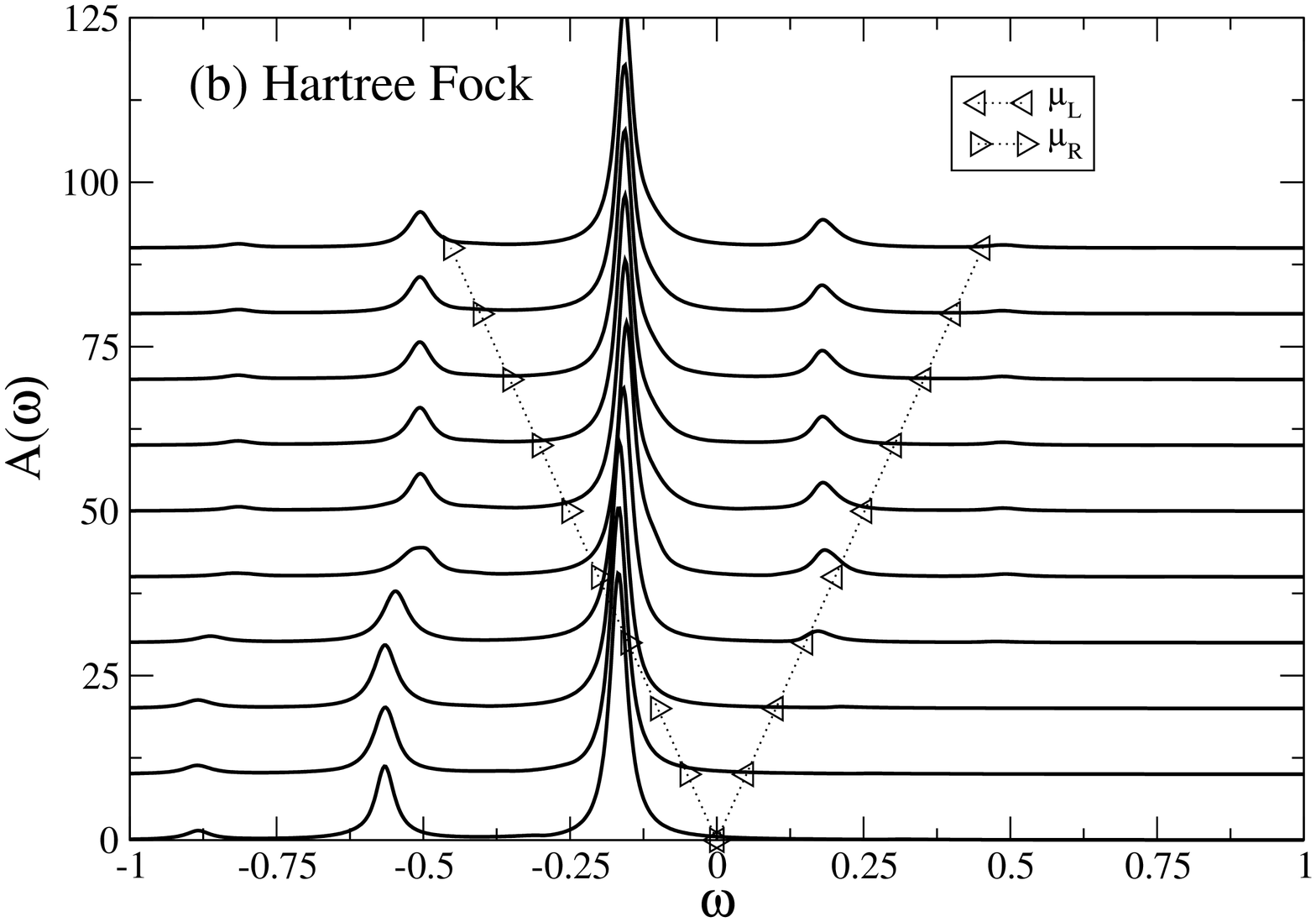}
  \caption{Non-equilibrium spectral function $A(\omega)$ for the 
    resonant transport regime for different applied biases for (a) the
    Fock diagram only, and (b) the Fock and Hartree diagrams. The
    curves are shifted vertically (+10 in $y$-axis) for clarity, with
    the values of the left and right chemical potentials given by the
    triangular symbols, here we have a symmetric potential drop.
    Adding the Hartree diagram breaks the electron-hole symmetry.
    The other parameters are the same as for Figure \ref{fig:specdens_NE_offres_F}
    except for $\varepsilon_0=+0$ and $\eta_V=0.5$.}
    \label{fig:specdens_NE_res_F}
\end{figure}

The spectral functions for the Fock diagram are shown in figure
\ref{fig:specdens_NE_res_F}(a).  As we have no Hartree term, the spectral
function is electron-hole symmetric at equilibrium.  Moreover, as we
have chosen a symmetric potential drop ($\eta = 0.5$) the spectral
functions stay symmetric for all applied biases.  Constraining the
symmetry in this way allows us to concentrate on the effects of
increasing the bias, and we can see that the most important
modifications of the spectral function are in the width of the central
and satellite peaks.  The vibron side-band peaks show only a very
small change with increasing applied bias.

Figure \ref{fig:specdens_NE_res_F}(b) shows the spectral function when
we add in the Hartree term while keeping all other variables
unchanged.  We note immediately that the electron-hole symmetry is
broken, even at equilibrium.  In addition the main peak is shifted
towards negative energies with respect to the Fock-only calculation,
and the right-hand vibron side-band peak is completely
suppressed---the whole spectral function takes on the qualitative
appearance of a spectral function in the hole-dominated regime
(compare with Figure \ref{fig:specdens_equi_offres}).  On increasing
the bias, the width and shape of the left-hand vibron side-band peaks
is varied, and for higher biases new vibron side-band peaks appear
above the main peak, until at high biases the spectral function is
very similar in shape to that for the off-resonant spectral function
(Figure \ref{fig:specdens_NE_offres_F}).

\subsection{A comparison with inelastic scattering techniques}
\label{sec:comp-with-inelastic-scattering-technique}

In this section, we will check the validity of the self-consistent
Born approximation (SCBA), i.e.  self-consistent calculations using
only the Hartree and Fock diagrams, versus another method which gives
more exact results as far as \elvib coupling is concerned.  Since NEGF
is a many-body perturbation expansion theory, by definition it does
not contain all diagrams even though self-consistency allows us to
achieve a partial resummation of a subclass of diagrams.

As an alternative to NEGF, one can calculate the transport properties using an
extension of conventional scattering theory to include the interaction
of incoming single-particle states with some bosonic degrees of 
freedom inside the central region of interest \cite{Bonca:1995,Ness:2001,
Ness:2006}.

This technique, termed the multi-channel inelastic scattering 
technique (MCIST) \cite{Ness:2001, Ness:2006}, has the
advantage of being an exactly solvable problem, even in the presence
of many non-interacting electronic states coupled to many vibration 
modes in the central region \cite{Ness:2001}.  
MCIST is based on many-body perturbation theory for polaron and it is
exactly solvable in the sense that MCIST is treating the \elvib 
coupling to all orders.  
In the language of polaron theory, MCIST contains all the diagrams
corresponding to the \elvib interaction (in the corresponding transport
regime).  To be more precise, MCIST contains all orders of crossing and
non-crossing diagrams in terms of the bare vibron propagator
\cite{Ness:2006,Ciuchi_S:1997,Cini_M:1988,Smondyrev:1986}. 
In conventional polaron theory for one electron, there are no diagrams 
with electron-hole loops in them \cite{Smondyrev:1986}.

However, MCIST is a single-particle scattering technique and treats the
statistics of the Fermi seas of the left and right leads only in an 
approximate manner.
In the language of NEGF, this means that the results given by MCIST are
only valid for a specific transport regime (as we will see below).

In the case of the SSSM model, the retarded \GF of the central region
has the usual form
\begin{equation}
  \label{eq:Gr_mcist}
  G^r(\omega)=[\ g^r_0(\omega)^{-1}-\Sigma^r_{\rm leads}(\omega)
  -\Sigma^r_{\evib}(\omega)]^{-1} \ .
\end{equation}
Within MCIST, the retarded \elvib self-energy containing all orders
of the \elvib coupling is expressed as a continued-fraction as 
shown analytically in Refs.~[\onlinecite{Ness:2006}] and [\onlinecite{Cini_M:1988}]:
\begin{equation}
  \label{eq:SEr_mcist}
  \begin{split}
   & \Sigma^r_{\evib}(\omega)  = \\
  &  \cfrac{\gamma_0^2}{
      G^r_0(\omega-\omega_0)^{-1}-
      \cfrac{2\gamma_0^2}{
        G^r_0(\omega-2\omega_0)^{-1}-
        \cfrac{3\gamma_0^2}{
          G^r_0(\omega-3\omega_0)^{-1}- ...}}} 
  \end{split}
\end{equation}
where we recall that $G^r_0$ is the retarded GF of the central
region connected to the leads, i.e. 
$G^r_0(\omega)=[\ g^r_0(\omega)^{-1}-\Sigma^r_{\rm leads}(\omega)]^{-1}$,
without \elvib coupling.

As show in Ref.~[\onlinecite{Ness:2006}], the lowest Born approximation,
fully consistent with SCBA (Hartree-Fock calculations as shown
above), is given in MCIST by an \elvib self-energy
equivalent to Eq.~(\ref{eq:SEr_mcist}) but where the integer $n$
factors at each level of the continued fraction are all replaced by
the integer $n=1$.

Obviously, this approximate substitution is good enough in the
(very) weak \elvib coupling for which only the first level of the 
continued fraction contributes the most \cite{Ness:2006}.

Below we compare the spectral functions obtained by exact MCIST 
calculations, by MCIST approximated to the SCBA (Hartree-Fock)
level, and to NEGF-SCBA calculations.

\begin{figure}
  \includegraphics[width=8cm]{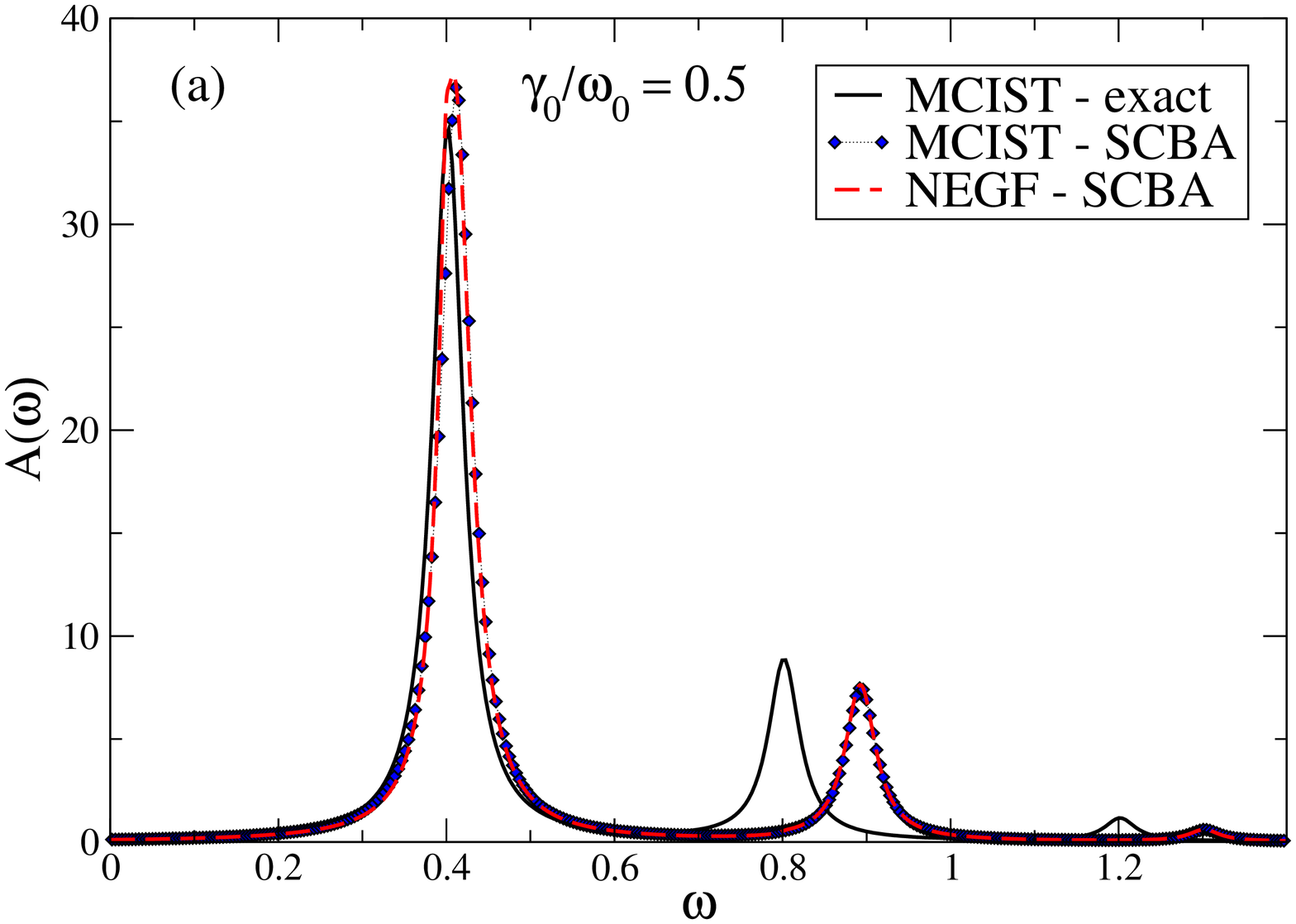}
  \\
  \includegraphics[width=8cm]{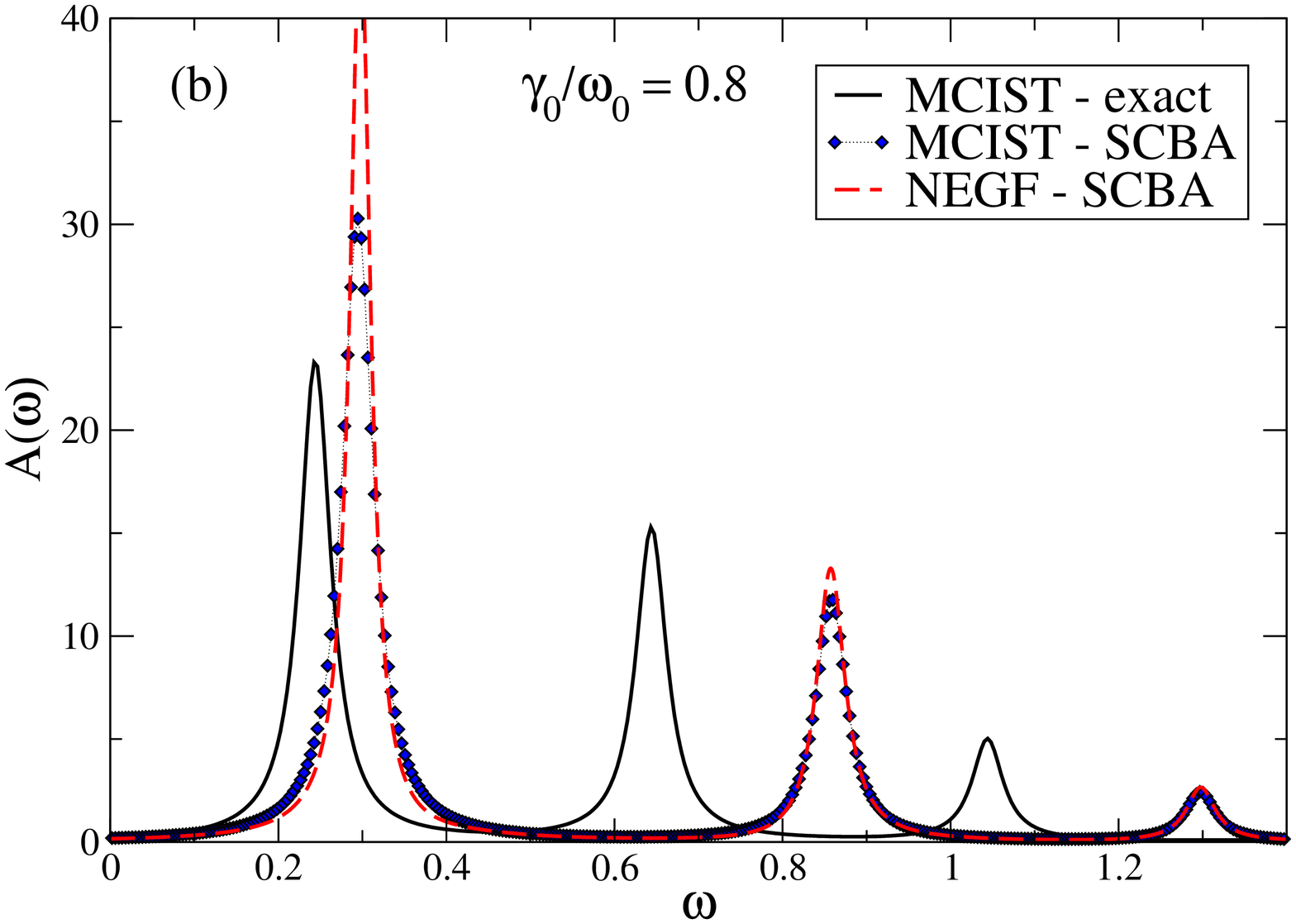}
  \caption{(Color online) Equilibrium off-resonant spectral functions
    calculated with the MCIST technique and NEGF-SCBA for the
    off-resonant regime.  The upper panel shows results for weak
    \elvib coupling ($\gamma_0/\omega_0 = 0.5$), while the lower panel
    shows results for strong \elvib coupling ($\gamma_0/\omega_0 =
    0.8$).  MCIST calculations give exact results in this case, and
    can also be performed at the same level of approximation as SCBA
    (see main text for details).  MCIST-SCBA and NEGF-SCBA are
    virtually identical, but show large discrepancies for the peak
    positions (main peak and more especially for the vibron side-band
    peaks) in comparison to the exact results.  The parameters are
    $\varepsilon_0=+0.5$ (electron transport), $\gamma_0=0.2$ (upper
    panel) and $\gamma_0=0.32$ (lower panel), $\omega_0=0.4,
    t_{0L,R}=0.15$.}
    \label{fig:specdens_MCISTversusSCBA}
\end{figure}

The spectral functions obtained from MCIST are shown in figure
\ref{fig:specdens_MCISTversusSCBA} for the weak/intermediate \elvib
coupling regime ($\gamma_0/\omega_0 = 0.5$) and for the strong \elvib
coupling regime ($\gamma_0/\omega_0 = 0.8$).  The overall lineshapes
correspond to a main peak with vibron side-band peaks located only
above the main peak. These are typical results fully consistent with
an off-resonant transport regime situation.  Furthermore, the spectral
functions obtained with BA-based approximation (MCIST-SCBA and
NEGF-SCBA) are virtually identical, especially in the intermediate
($\gamma_0/\omega_0 = 0.5$) to very weak (not shown here) \elvib
coupling regime.  For strong \elvib coupling, one obtains the same
peak positions, however the amplitude of the peaks (especially the
main peak) is slightly different.

Since MCIST calculations will always give similar lineshapes
independent of the value of $\varepsilon_0$, one can conclude that
MCIST calculations are only valid for the off-resonant transport
regime at and near equilibrium.  MCIST is not able to reproduce the
spectral features of the resonant transport regime (i.e. vibron
side-band peaks on both sides of the main peak).  This is essentially
due to the fact that in MCIST, one does not take properly into account
the statistics of the Fermi seas of the left and right leads.  See for
example Eqs.~(\ref{eq:Gr_mcist}) and (\ref{eq:SEr_mcist}), there are no
leads' Fermi distributions in the retarded component of the leads'
self-energies $\Sigma^r_{L,R}$.

Now, comparing BA-based calculations with the exact MCIST
calculations, one can see from figure
\ref{fig:specdens_MCISTversusSCBA} that the BA-based calculations give
the wrong polaron shift, i.e. the normalized position of the main peak
$\varepsilon_0$, especially in the strong \elvib coupling
regime. Furthermore BA-based calculations also give the wrong energy
separation between the main peak and the first vibron side-band
peak. This energy difference should be equal to the vibron energy
$\omega_0$, as it is given by exact MCIST calculations.  Note that the
limits of BA-based calculations were also been studied by Lee {\it et al.}
in a somewhat different context in Ref.~[\onlinecite{Lee:2009}].

In conclusion, this means that Hartree-Fock (or BA) based calculations
for \elvib interaction are only valid for weak coupling, as can be
expected from a perturbation-expansion based theory.  Hence one needs
to include higher-order diagrams in the \elvib self-energies to go
beyond the commonly-used self-consistent Born approximation
(Hartree-Fock) in order to obtain correct results for a wide range of
parameters.  The effects of the higher-order diagrams (here
second-order-DX and DPH diagrams) are explored in detail in the
following sections.

Additionally, although MCIST calculations are only valid in the
off-resonant transport regime at and near equilibrium, they include
all possible higher-order diagrams (with bare vibron propagator) 
and hence can be used as a reference for any perturbation-expansion-based 
NEGF calculations performed at equilibrium or in the quasi-equilibrium
regime.



\subsection{Vertex corrections and polarization effects
to the spectral functions}
\label{sec:vert-corr-spectr}

In this section, we present results for
the spectral functions when the second-order diagrams (see
Figure \ref{fig:DX-DPH}) are included in the calculations of the
Green's functions.  The reader can find more information about the mathematical expressions
for the self-energies corresponding to the second-order diagrams 
in Appendix \ref{app:higherorderdiagrams}.

These diagrams fall into two types---the double-exchange DX diagram,
corresponding to vertex corrections, and the dressed vibron diagram, which
includes a single electron-hole bubble, renormalizing the
vibron propagator and hence giving rise to polarization effects.

We have used three different levels of approximation to calculate
these Green's functions: Firstly, calculations with no
self-consistency---the Green's functions are simply calculated using
the diagrams in Figure \ref{fig:FockHartree} and Figure
\ref{fig:DX-DPH} using the bare propagator $G_0$ as the electron
Green's function.  In our model, $G_0$ is the Green's function of the
central region connected to the leads with no \elvib interactions.
This is a first-order perturbation expansion for which
$\Sigma_\evib^{H,F,DX,\ {\rm and/or}\ DP}[G_0]$.  We use the
abbreviations BA (Born approximation for non self-consistent Hartree
and Fock diagrams) and BA+DX (DX for double exchange) and BA+DX+DPH
(DPH for dressed vibron, for the $GW$-like diagram) in the following.

Secondly, we perform partly self-consistent calculations, where the
Green's functions are calculated with the first loop of self-consistent
calculations with the Hartree and Fock diagrams.  We use these Green's functions
as a starting point to calculate new, corrected, Green's functions
including the second-order diagrams $\Sigma_\evib^{\text{DX}}$ and/or
$\Sigma^{GW}_\evib[G^{SCBA}]$ in which the Green's functions are the corresponding SCBA
Green's functions. 

Finally, we perform fully self-consistent calculations, in which the
Green's functions are calculated in a self-consistent manner with the
first and second-order diagrams included within each iteration of the
self-consistency loop.

Our rationale for choosing to do calculations in this manner is as
follows.  We are able to test the different levels of
approximation and see very precisely the effects of vertex corrections
and polarization are on both the bare Green's functions $G_0$ and the
SCBA-level Green's functions.  We will also show in section
\ref{sec:2ndorder_off_res_equi} that by using the SCBA Green's
functions rather than $G_0$ as a starting point, one achieve a
better convergence in the calculations.  
In particular we will show later on in Section \ref{sec:2ndorder_off_res_equi}
that in some cases, for example the off-resonant transport
regime when $\gamma_0/\omega_0 \gtrsim 0.6-0.7$), the use of the bare
$G_0$ Green's functions as a starting point for a fully self-consistent
calculation with second-order diagrams actually gives unphysical results.
Additionally, fully self-consistent calculations are extremely
computationally intensive, and hence it is both interesting and useful
to explore the range of parameters for which the second-order-diagram
corrections to the SCBA calculations give a sufficiently accurate
description of the Green's functions in comparison to the fully
self-consistent calculations.

\subsubsection{Off-resonant regime at equilibrium}
\label{sec:2ndorder_off_res_equi}

\begin{figure}
  \includegraphics[width=8cm]{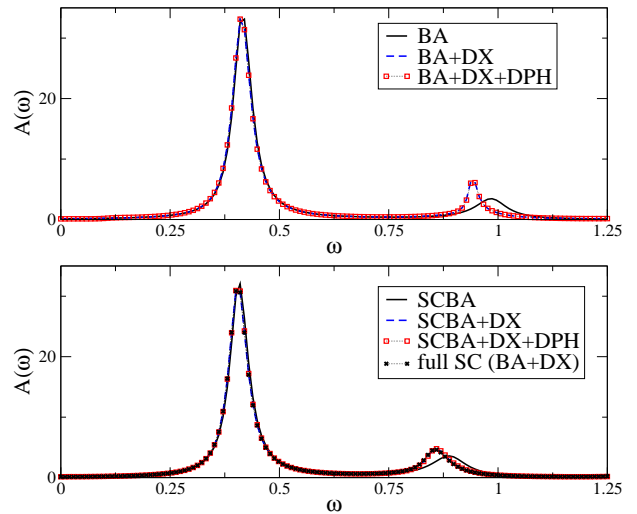}
  \caption{(Color online) Equilibrium spectral function for the off-resonant
    electron transport regime in the weak/intermediate \elvib coupling.
    Non self-consistent (top panel) and self-consistent (bottom panel) 
    calculations were performed with first-order Hartree and Fock-like
    (BA, solid lines), plus second-order DX (dashed lines), plus DPH
    (squares) \elvib self-energy diagrams.  
    The other parameters are $\varepsilon_0=+0.5$, $\gamma_0=0.2,
    \omega_0=0.4, t_{0L,R}=0.15, \eta=0.025$.
    Self-consistent calculations give better spectral functions, with
    vibron side-band peak closer to the main peak. The second-order DX diagram
    narrows the side-band peaks and bring them even closer to the main peak
    as expected. The second-order DPH diagram does not contribute much in the 
    off-resonant regime.
    In this regime and for small \elvib coupling, second-order corrections
    to SCBA calculations are a good approximation to corresponding full  
    self-consistent calculations.}
    \label{fig:specdens_eq_ofres_HFDXDPH}
\end{figure}

Figure \ref{fig:specdens_eq_ofres_HFDXDPH} shows the spectral functions of
the off-resonant transport regime at equilibrium, calculated for 
weak/intermediate \elvib coupling ($\gamma_0/\omega_0=0.50$) and for
different diagrams and levels of self-consistency.

As already mentioned in sections \ref{sec:at-equilibrium} and
\ref{sec:NonEq-SpecFns},
the energy separation
between the main peak and the first vibron side-band peak (which should be
equal to the vibron energy, here $\omega_0=0.40$) is not well reproduced
by BA-based (Hartree-Fock based) calculations and is much larger than
$\omega_0$.
The self-consistency introduced in the calculations give a marginally
smaller energy separation.

The effects of the second-order DX diagram are firstly  to bring the vibron
side-band peak closer to the main peak, hence giving an energy separation
closer to the exact value; and secondly, one observes a strong narrowing and
a larger amplitude of the vibron side-band peak in both self-consistent
and non-self-consistent calculations.
Both these effects thus qualitatively modify the spectral functions towards
better agreement with the exact results as shown 
in Section \ref{sec:comp-with-inelastic-scattering-technique}.

It is, however, worth mentioning that including the DX diagram in the calculation
does not greatly affect the position of the main peak, a result which may
be understood from the fact that the dynamical polaron shift is a quantity 
difficult to obtain exactly from a perturbation expansion theory beyond weak
coupling \cite{Ness:2006,Smondyrev:1986}.

The effects of the second-order DPH diagram in the spectral function
are virtually nil for the case of weak/intermediate coupling and the
off-resonant regime.  This might not be that surprising since the DPH
diagram corresponds to a Fock-like diagram with a renormalised vibron
propagator. The renormalization of the vibron is due to a single
electron-hole bubble. However in the off-resonant transport regime,
the spectral function 
is almost empty (in the case of electron
transport) or almost full (in the case of hole transport) which
implies that there are not many electron-hole excitations available in
this transport regime. Hence the polarization (i.e. the contribution
of the electron-hole bubble) is very small, subsequently giving very
small values for the DPH self-energy.
As an example, we have checked our numerical values in the case shown in
figure~\ref{fig:specdens_eq_ofres_HFDXDPH} and we have found, as expected,
that the maximum values of $\Sigma_\evib^{{\rm DPH},r}$ are 30 to 50 times
smaller than the maximum values of $\Sigma_\evib^{{\rm DX},r}$.

Furthermore, in the off-resonant regime and for weak-ish \elvib
coupling, it seems that the fully self-consistent (full SC (BA+DX)
curve in figure~\ref{fig:specdens_eq_ofres_HFDXDPH}) results $G_{\rm
  full\ SC}^r(\omega)$ are well approximated by the results given by a
second-order correction to a self-consistent Hartree-Fock (SCBA)
calculations (SCBA+DX curve in
figure~\ref{fig:specdens_eq_ofres_HFDXDPH}) $G_{\rm partial\
  SC}^r(\omega)=[G_{\rm SCBA}^r(\omega)^{-1} - \Sigma^{{\rm
    DX},r}[G_{\rm SCBA}]\ ]^{-1}$.  This is an interesting result
as it implies that the physical properties of the system, at least in
these conditions, could be well described by a second-order correction of
the lowest-order SC calculations, without the need to perform a fully
self-consistent calculation up to the second-order.

\begin{figure}
  \includegraphics[width=8cm]{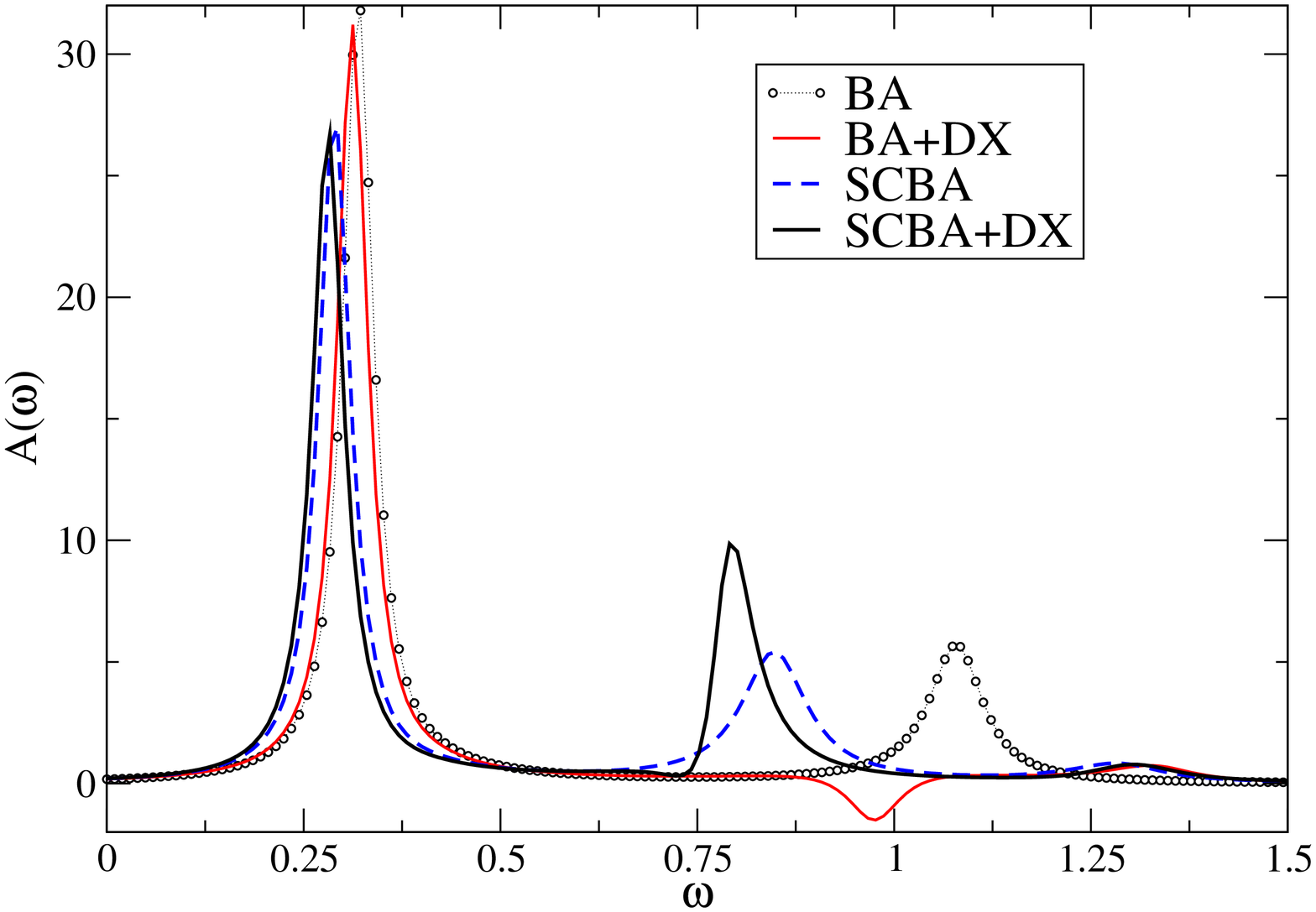}
   \vspace{2mm}
  \\
  \includegraphics[width=8cm]{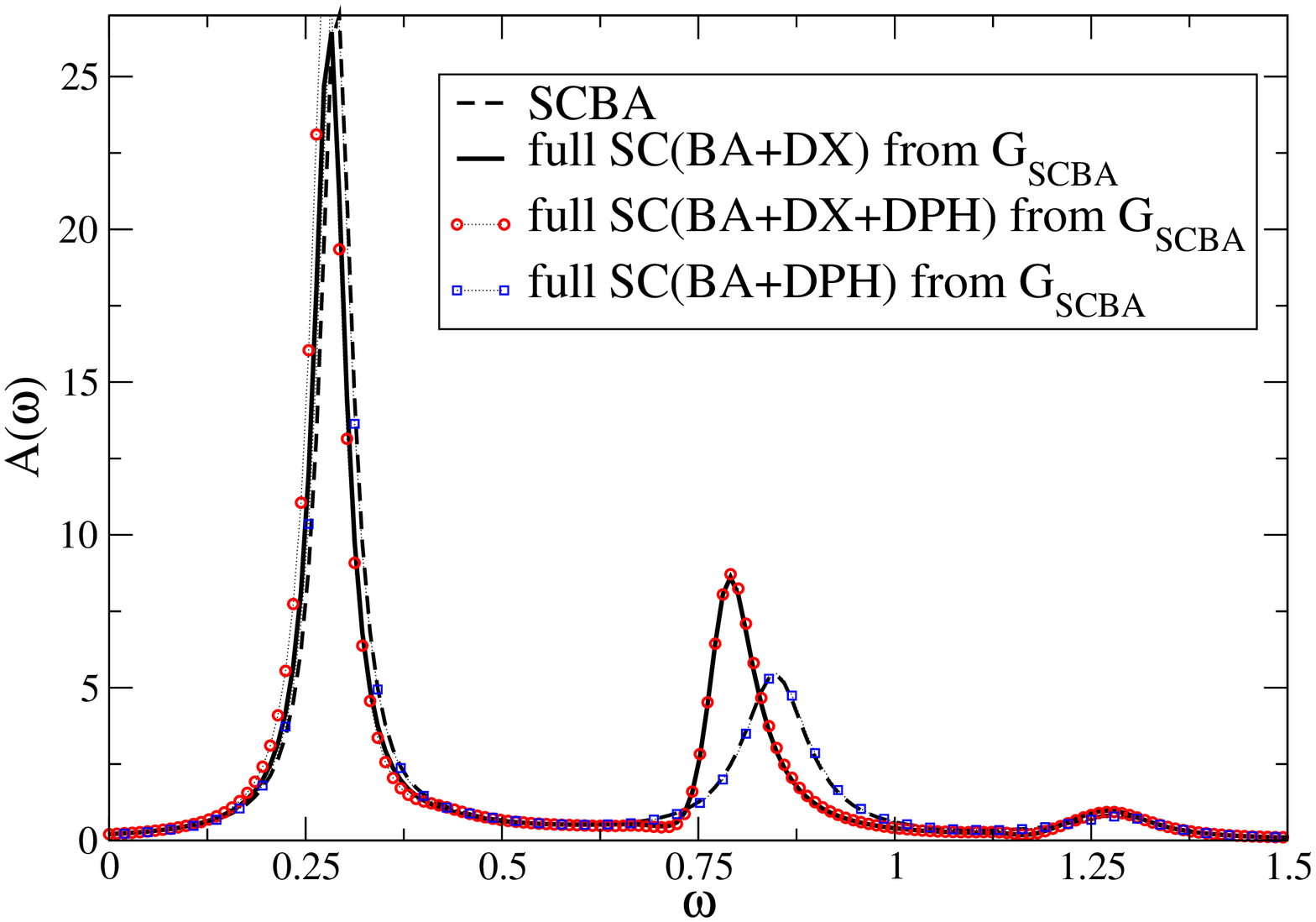}
  \caption{Equilibrium spectral function for the 
    off-resonant electron transport regime and strong \elvib coupling.
    Calculations were performed with different diagrams:
    first-order Hartree and Fock-like (BA)
    and second-order (DX, DPH) \elvib self-energy diagrams.
    (Top panel):
    Calculations were done with no self-consistency (BA,BA+DX), partial (SCBA+DX) 
    and full self-consistency (SCBA).
    (Bottom panel): full self-consistent SC(BA+...) calculations.     
    The other parameters are $\varepsilon_0=+0.5$, 
    $\gamma_0=0.32, \omega_0=0.4, t_{0L,R}=0.15, \eta=0.025$.
    The calculations show that no self-consistency gives poor results in comparison
    to exact calculations given in Figure \ref{fig:specdens_MCISTversusSCBA}.
    A closer spectral function to the exact result is obtained from SC(BA+DX) 
    calculation. The second-order DPH diagram do not contribute much in the 
    off-resonant regime.
    See main text for a detailed analysis of the spectral functions.}
    \label{fig:specdens_eq_ofres__strg_HFDXDPH}
\end{figure}

Now we turn to the analysis of results obtained for stronger \elvib
coupling.  Examples of such calculations are given in
figure~\ref{fig:specdens_eq_ofres__strg_HFDXDPH} for
$\gamma_0/\omega_0=0.80$.  In the strong \elvib coupling regime, one
obtains qualitatively the same contributions of the second-order
diagrams as explained above for the weak coupling regime.  The DPH
diagram does not have a large role in the off-resonant transport
regime, although slightly affecting the width of the main peak.  And
the DX diagram shifts the vibron side-band peak towards the main peak
(and hence towards the exact results) as well as narrowing the peak
width and increasing the peak amplitude.  These effects are amplified
in Figure~\ref{fig:specdens_eq_ofres__strg_HFDXDPH} because the \elvib
coupling constant $\gamma_0$ is bigger than in
Figure~\ref{fig:specdens_eq_ofres_HFDXDPH}.  Furthermore, for strong
\elvib coupling, the DX diagram modifies the energy position of the
main peak and seems to give a slightly better polaron shift.

There is however another interesting effect observed from such a set
of calculations, which can be seen in the upper panel of
Figure~\ref{fig:specdens_eq_ofres__strg_HFDXDPH}.  The first iteration
of a fully self-consistent calculation (including the Hartree, Fock,
and DX diagrams) starting with the non-interacting GFs $G_0$ gives non
physical results, i.e. negative values of the spectral function (see
curve BA+DX in figure~\ref{fig:specdens_eq_ofres__strg_HFDXDPH}), and
such an unphysical behaviour does not self-correct in the following
iterations.  This a well known problem, which has already been
encountered in the past by several authors in the context of \elvib
interaction (for example in Ref~[\onlinecite{Kral:1997}]) and also in
the context of electron-electron interaction when considering
topologically equivalent diagrams \cite{Shirley:1993a,Shirley:1996}
\endnote{Ulf von Barth, private communication}.
In Ref.~[\onlinecite{Kral:1997}], a somewhat different approach than
ours was used.  It is based on a linked cluster expansion for the 
non-equilibrium steady-state regime, and negative densities of states
were obtained when including second-order diagrams, and in some cases
even higher-order cluster approximations did not seem to give a
convergent solution at intermediate \elvib strength and in the
presence of the Fermi seas.

However, our calculations reveal that it is possible to solve such a problem by 
starting the fully self-consistent calculations (up to second order) from a 
different starting point, namely by starting from the $G_{\rm SCBA}$ Green's
functions (i.e. the GFs obtained from a fully SC calculation including only
the lowest-order diagrams). This is shown by the curve (SCBA+DX) in the upper
panel of figure~\ref{fig:specdens_eq_ofres__strg_HFDXDPH} and by the lower
panel in which all the fully converged self-consistent results are shown.

For the moment, we do not have a full physical explanation of the reason
why starting from a SCBA calculations is better to achieve full self-consistency
with higher-order diagram than Hartree-Fock, apart from the simple fact that an
Hartree-Fock calculation is probably closer to the true interacting solution
than the non-interacting solution.

To conclude this  section, we can say that in the off-resonant
regime at equilibrium, the second-order DX diagram dominates over the second
order DPH diagram.

\subsubsection{Resonant regime at equilibrium}
\label{sec:2ndorder_res_equi}

In this section, we present calculations for the resonant regime at
equilibrium. Though we have shown, in sections \ref{sec:SpectralFunctionsResonant} 
and \ref{NonEq-SpecFns-Resonant}, the importance of the Hartree diagram,
we will consider below results obtained without the Hartree diagram. 
The calculations were performed with the Fock and second-order DX and/or
DPH diagrams which conserve (at numerical accuracy) the electron-hole
symmetry of the system.

The reasons why we have choosen to perform this model calculation
are twofold: 
first, even for a single electronic level, we expect to have the 
maximum possible electron-hole excitations available when the 
spectral functions are electron-hole symmetric.
Hence we expect the polarization effects be to more pronounced in 
a system with electron-hole symmetry.
Second, the electron-hole symmetric model permits us to emphasize the 
competitive effects between the DX and DPH diagrams as will be shown
below.  

\begin{figure}
  \includegraphics[width=8cm]{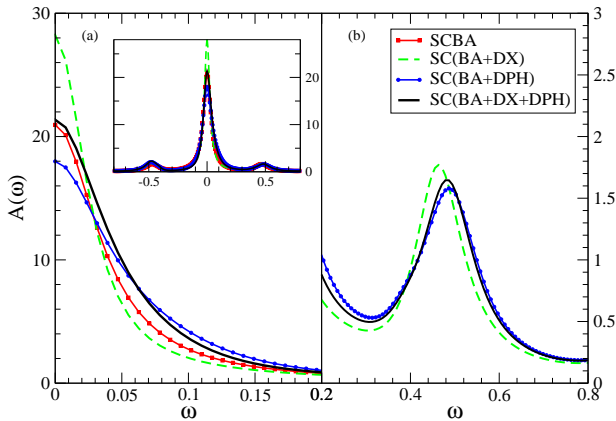}
  \caption{(color online). Equilibrium spectral function for the 
    resonant transport regime and weak/intermediate \elvib coupling.
    Calculations were performed with diagrams:
    BA (Hartree-Fock) and second-order DX, DPH \elvib diagrams.
    The other parameters are
    $\varepsilon_0=+0.0$ (resonant transport), 
    $\gamma_0=0.24, \omega_0=0.4, t_{0L,R}=0.2, \eta=0.03$.
    The top inset shows the whole spectral functions, while the left
    (a) and right (b) parts are zooms, for positive $\omega$, of the
    central peak and first vibron side-band peak respectively.
    The second-order diagrams have the following effects on the SCBA
    spectral functions: DPH broadens the central and side-band peaks as
    well as lowers their height, in opposition to DX which narrows the
    peaks and brings the side-band peaks slightly closer to the center,
    with a strong and unphysical increase of the amplitude of the central
    peak.
    Calculations with both DX and DPH give in appearance a broadening of the 
    central peak, which however recovers the correct height necessary to 
    conserve the Fermi-liquid property of the electron-hole symmetric 
    system (see Section \ref{sec:linear_reponse} for more detail). }
    \label{fig:specdens_eq_res__intermed_HFDXDPH}
\end{figure}

Figure \ref{fig:specdens_eq_res__intermed_HFDXDPH} shows the equilibrium
spectral functions of the resonant regime (at weak/intermediate \elvib
coupling) obtained from different self-consistent calculations including
first and second-order diagrams.

On one hand, the second-order DPH diagram corresponds to a partial 
dressing of the vibron propagator by one electron-hole bubble. 
It gives an extra lifetime in the retarded/advanced electron Green's 
functions in comparison to Hartree-Fock (SCBA) calculations.
Hence the main effect of the DPH diagram is to introduce an extra broadening 
of the peaks in the spectral functions.
Since the DPH diagram is one of the so-called conserving approximations
\cite{Baym:1962,Kadanoff:1962}, the broadening of the peaks leads to a reduction 
of their height to keep globally the same total spectral weight. 
These effects can be seen on the SC(BA+DPH) curve 
in Figure \ref{fig:specdens_eq_res__intermed_HFDXDPH}.

On the other hand, the second-order DX diagram, which is a 
conserving approximation, has an opposite effect: a strong narrowing of the
peaks (especially of the central peak) accompanied with an increase of their
height, as can seen on the SC(BA+DX) curve in 
Figure \ref{fig:specdens_eq_res__intermed_HFDXDPH}.

Now it is interesting to see what happens when the calculations are
performed with both second-order DX and DPH diagrams. This is shown on
the SC(BA+DX+DPH) curve in Figure
\ref{fig:specdens_eq_res__intermed_HFDXDPH}: We obtain a hybrid
behaviour, in the sense that in appearance the central peak is
broadened in comparison to SCBA calculations. However, the height of
the peak at the Fermi level is conserved (up to numerical accuracy).
This is a very important result which proves that for the model
calculation of an electron-hole symmetric system, one has to include
both the DX and DPH second-order diagrams in order to conserve the expected
Fermi-liquid properties of the system. In this regime both the DX and DPH
second-order diagrams play an equally important role which determines
the linear response properties of the system as shown below in Section
\ref{sec:linear_reponse}.

By comparison, in a diagrammatic treatment of the
\textit{electron-electron} interaction on the electron propagator, the
situation is often different: the electron-hole bubble diagram that
appears here in the DPH contribution to the propagator is large,
especially in highly polarizable metallic and open-shell systems where
electron-hole pairs may be created with low energy cost, because the
Coulomb interaction operates at all energy scales.  In that case,
summing the bubble diagrams to infinite order as is done in Hedin's
$GW$ approximation \cite{Hedin:1965,Hedin:1969} is much more important 
than including the second-order exchange diagram.  
The key difference for the electron-phonon interaction is that the vibron 
frequency $\omega_0$ imposes a restricted energy scale on the interaction, 
reducing the importance of the bubble diagrams, and correspondingly 
increasing the importance of the second-order exchange diagram.

\subsubsection{Off-resonant regime at finite bias}
\label{sec:2ndorder_off_res_NE}

\begin{figure}
  \includegraphics[width=8cm]{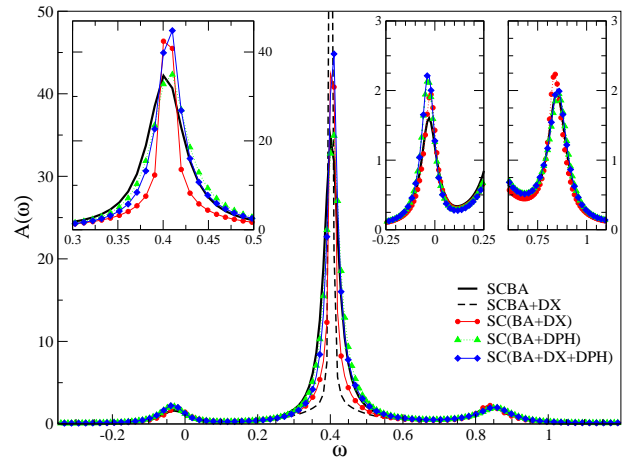}
  \caption{(color online). Non-equilibrium spectral function 
    (bias $V=0.45>\omega_0$) for the 
    off-resonant transport regime for the weak/intermediate \elvib coupling.
    Calculations were performed self-consistently with first-order Hartree and 
    Fock-like diagrams (SCBA curve), and with second-order DX and/or DPH diagrams
    (SC(BA+DX,+DPH,+DX+DPH) curves). Second-order DX correction to SCBA is also 
    shown (SCBA+DX curve).
    The top-left inset shows a zoom on the central peak around $\omega\sim 0.4$,
    and the top-right insets show a zoom on the vibron side-band peaks around 
    $\omega\sim 0.0$ and $\omega\sim 0.8$.
    The other parameters are
    $\varepsilon_0=+0.5$ ,
    $\gamma_0=0.2, \omega_0=0.4, t_{0L,R}=0.15, \eta=0.025, \eta_V=1$.
    The second-order DX diagram narrows the peaks with a slight shift of the side-band
    peaks towards the central peak. The narrowing is too strong in the case of
    partially self-consistent calculations (SCBA+DX). The second-order DPH diagram 
    broadens the peak, but not as much as in the resonant case. Full self-consistent
    calculations including both second-order diagrams result in an intermediate 
    behaviour for the modifications of the spectral functions.}
    \label{fig:specdens_NE_ofres_HFDXDPH}
\end{figure}

Figure \ref{fig:specdens_NE_ofres_HFDXDPH} shows the spectral
functions of the off-resonant transport regime at finite bias.  The
calculations have been performed for weak/intermediate \elvib coupling
($\gamma_0/\omega_0=0.50$) and for different diagrams and levels of
self-consistency.  We have considered the case for which real
excitations of vibrons are possible $V=0.45 >\omega_0$.

The interpretation of the results is not as straightforward as in the
equilibrium case, because non equilibrium effects are sometimes
counter intuitive.  However, the overall shapes of spectral
functions are quite similar to that obtained within Hartree-Fock based
calculations (see Section \ref{NonEq-SpecFns-OffRes}) in the sense
that they present a central peak with vibron side-band peaks on both
sides.

Similarly to the resonant regime at equilibrium, the main effects of the 
second-order DX diagram is to narrow the width of all peaks, and to shift
slightly the side-band peaks towards the main peak.

As mentioned in the previous section, the main effect of the second-order
DPH diagram is to broaden the peaks, in opposition to the effects of the
DX diagram. However here, the broadening in the non-equilibrium
condition appears less important than in the case of the resonant regime
at equilibrium. 
Hence full self-consistent calculations performed with both DX and DPH 
second-order diagrams give a narrowing of the peaks with a corresponding 
increase of their amplitude in comparison to SCBA calculations.

Finally, out of equilibrium, it can be seen that the results given by
a full self-consistent calculation (curve SC(BA+DX) in Figure
\ref{fig:specdens_NE_ofres_HFDXDPH}) are strongly different from a
second-order correction to an Hartree-Fock calculation (SCBA+DX curve
in \ref{fig:specdens_NE_ofres_HFDXDPH}) which gives an excessively
narrowed central peak.

This means that in the weak/intermediate \elvib coupling, second-order
corrections to a SCBA calculations are only good enough at equilibrium,
however at non-equilibrium full self-consistency needs to be performed
with all diagrams of the same order.

\subsubsection{Resonant regime at finite bias}
\label{sec:2ndorder_res_NE}

As we have already shown in the first-order \elvib diagrammatic
calculations, the spectral functions for the off-resonant and resonant
regime at non-equilibrium are qualitatively similar, in the sense that
they present a central peak with vibron side-band peaks on both
sides. One can compare for example the spectral functions obtained for
Hartree-Fock-like calculations at non-equilibrium shown in Figure
\ref{fig:specdens_NE_offres_F} and Figure \ref{fig:specdens_NE_res_F}.

Hence, and again on a qualitative level, the effects of the second-order
DX and DPH diagrams on the resonant case at finite bias are similar to what has
been obtained for the off-resonant non-equilibrium case described in the previous
section.  The effects of these higher-order diagrams on 
the full non-equilibrium transport properties for the different transport regimes
will be presented in a forthcoming paper.

However before turning the discussion to the linear-response properties 
of the system at and near equilibrium, we would like to comment on a specific 
aspect of the effects of higher-order diagrams.

The narrowing of the vibron side-band peaks and of the main central
peak due to higher-order (DX) diagrams was also obtained by other
authors (see for example Ref.~[\onlinecite{Zazunov:2007}]).  In this
paper, a different non-equilibrium approach was used. It consists of
starting with an electron dressed by a vibron (a polaron) in the isolated
central region, then using perturbation expansion theory (with partial
resummation) in terms of the coupling of the central region to the
non-equilibrium left and right leads.  However, the results for the
spectral functions in Ref.~[\onlinecite{Zazunov:2007}] were only given
for the resonant transport regime at equilibrium and all calculations
were performed without the Hartree-like diagram. Though we have
already shown that such a diagram plays a crucial role in the spectral
properties of the system in the resonant regime at and out of
equilibrium.

\subsection{Linear response transport properties}
\label{sec:linear_reponse}

We now briefly discuss the effects of \elvib interaction taken at different 
levels of the diagrammatic expansion on the linear-response transport properties
of the single-molecule nanojunction.
Before doing so, we explain how to derive the linear conductance from the value
of the spectral functions calculated at equilibrium.

In our model, there is a direct proportionality between the left and right leads' 
self-energies $\Sigma^{r}_{L,R}$ because, the central region is
coupled to the leads via the single hopping matrix element and we have chosen
identical leads.
Then the current (Eq.~(\ref{eq:Expectation_value_current})) can be recast as
follows (for the details of the derivation, see for example 
Refs.[\onlinecite{MeirWingreen:1992,Ness:2006}]):
\begin{equation}
\label{eq:current_intermsof_ImGr}
I=-4e/h \int {\rm d}\omega \left( f_L(\omega)-f_R(\omega)\right) \Gamma(\omega) \Im m [G^r(\omega)],
\end{equation}
where $\Gamma=\Gamma_L\Gamma_R/(\Gamma_L+\Gamma_R)$ and $\Gamma_\alpha=-2\Im m[\Sigma^r_\alpha]$.

The linear conductance
\begin{equation}
G_{\rm lin} =\frac{dI}{dV}|_{V\rightarrow 0} \nonumber
\end{equation}
is obtained from
\begin{equation}
\label{eq:linear_conductance}
G_{\rm lin} / G_0 = -2\Gamma(\mu^{\rm eq}) \ \Im m [G^r(\mu^{\rm eq})]
= 2\Gamma(\mu^{\rm eq}) \ A(\mu^{\rm eq})
\end{equation}
where $G_0=2e^2/h$ is the quantum of conductance.

Table \ref{tab:linear_conduc} shows the different values for the linear
conductance $G_{\rm lin}$ obtained from the equilibrium spectral functions shown in 
figures \ref{fig:specdens_eq_ofres_HFDXDPH}, \ref{fig:specdens_eq_ofres__strg_HFDXDPH}      
and \ref{fig:specdens_eq_res__intermed_HFDXDPH}. 

\begin{table}\label{tab:linear_conduc} 
\begin{tabular}{|l|c|c|c|}
\hline 
$G_{\rm lin}/G_0$     & off-resonant  & off-resonant  & resonant  \\
              & weak \evib       & strong \evib  & {(e-h symmetric)}\\ 
              & Fig.~\ref{fig:specdens_eq_ofres_HFDXDPH}      
            & Fig.~\ref{fig:specdens_eq_ofres__strg_HFDXDPH}      
            & Fig.~\ref{fig:specdens_eq_res__intermed_HFDXDPH}     \\ \hline\hline 
SCBA             & 0.002654    & 0.004470    & 0.8379$^*$            \\ \hline
SCBA+DX         & 0.002666     & 0.004606    & -        \\ \hline
SCBA+DX+DPH         & 0.003186     & -         & -        \\ \hline 
SC(BA+DX)          & 0.002668     & 0.004606    & 1.1296        \\ \hline
SC(BA+DPH)        & -         & -        & 0.7194        \\ \hline
SC(BA+DX+DPH)        & -         & 0.004638     & 0.8545        \\ \hline\hline
MCIST             & 0.002621     & 0.004936     & n.a.        \\ \hline\hline
no \evib         & 0.002021     & 0.002021     & 1.        \\ \hline
\end{tabular} \newline 
\footnotesize{$^*$ in principle here $G_{\rm lin}$ should be $G_0$ but is not because we
are using a tiny but finite $\eta=0.030$ value. See appendix~\ref{app:numericalbits}
for detailed explanations.}
\caption{Linear conductance $G_{\rm lin}={dI}/{dV}|_{V\rightarrow 0}$
obtained from the value of the equilibrium spectral density at $\omega=\mu^{\rm eq}$.
$G_{\rm lin}$ is given in units of the quantum of conductance $G_0=2e^2/h$.
The equilibrium spectral functions are shown in 
Figs. ~\ref{fig:specdens_eq_ofres_HFDXDPH}, ~\ref{fig:specdens_eq_ofres__strg_HFDXDPH} and 
~\ref{fig:specdens_eq_res__intermed_HFDXDPH}) and were done for different
different levels of approximation as explained in the corresponding figure captions.} 
\end{table} 

For the off-resonant regime at the weak \elvib coupling, there are not
many differences in the linear conductance $G_{\rm lin}$ values
calculated for all the different diagrams, as one might expect from
the spectral function behaviour shown in
figure~\ref{fig:specdens_eq_ofres_HFDXDPH}.  This is especially true
when the main peak in the spectral function is located well away from
the Fermi level, i.e. $\vert\tilde\epsilon_0-\mu^{\rm eq}\vert \gg$
linewidth of the peaks.  For strong \elvib coupling, the differences
between different levels of approximations are more pronounced.  All
$G_{\rm lin}$ values are smaller than the exact MCIST result. This is
essentially due to the fact that perturbation expansion gives only an
approximate value of the polaron shift.  The differences from the
exact result are more important for SCBA-based calculations
(difference of $\sim 10 \%$ with exact result) and of course
calculations performed up to second order give a better polaron shift
and hence better values for the linear conductance ($\sim 5 \%$
difference only).

In the quasi-resonant case when $\vert\tilde\epsilon_0-\mu^{\rm
  eq}\vert\sim$ (1 to 2) linewidth, the effects described above will
be more pronounced, because the linear conductance is no longer given
by the extreme tail of the main peak crossing the Fermi level.

For the resonant case in the absence of the Hartree potential, the
SCBA-based calculations should give a perfect linear conductance. The
reason why $G_{\rm lin}\ne G_0$ is because the calculations were
performed with a tiny but finite value of $\eta$ as explained in
detail in Appendix~\ref{app:numericalbits}.  Hence one may say that,
for the chosen set of parameters, the value $G_{\rm lin}=0.838$
represents an upper bound for the linear conductance (the
corresponding numerical perfect conductance).  And as explained in
Appendix~\ref{app:numericalbits}, linear conductance values can only
be compared between calculations performed with the same value of
$\eta$, which is what we have done.

The linear conductance is strongly renormalized (decreased and
increased) when including only one of the two second-order diagrams
(DPH and DX respectively).  In contrast, it becomes close again to the
expected quantum of conductance when the calculations are performed
with both second-order diagrams.  This dependence of the conductance
is well understood from the behaviour of the spectral functions shown
in Figure \ref{fig:specdens_eq_res__intermed_HFDXDPH}.  SC(BA+DPH)
calculations introduce an extra broadening and a corresponding
decrease of amplitude of the peaks in the spectral functions, hence a
decrease of $G_{\rm lin}$.  However, the strong narrowing, with
increased amplitude, of the peaks due to SC(BA+DX) calculations leads
to an increase of $G_{\rm lin}$.  The SC(BA+DX) calculations give a
linear conductance larger than that obtained from SCBA, i.e.  $G_{\rm
  lin} > G_0$, which is an unphysical result for our non-degenerate
electronic level in the central part of the system.

This implies a strong constraint on the validity of the values of
$G_{\rm lin}$ obtained from a many-body perturbation expansion: in
order to conserve the Fermi-liquid properties (see
Appendix~\ref{app:numericalbits}) in the electron-hole symmetric
resonant regime, one has to perform the calculations by including all
diagrams order by order. Results obtained from partial resummation of
a subset of diagrams will probably give an incorrect linear
conductance.

Hence calculations performed by fully renormalizing the vibron
propagators \cite{Galperin:2004,Viljas:2005} or all crossing diagrams
\cite{Zazunov:2007}, while containing diagrams higher than second
order, will probably break important physical properties of the system
at $T=0$, because they lack important electronic processes, to
second-order the DX vertex correction-like diagram or DPH diagram
respectivelly.

In the resonant (or quasi-resonant) regime, polarisation effects and
vertex corrections play an equally important role in the electronic
structure and transport properties of \elvib interacting nanojunctions
which electron-hole symmetry.

\section{Conclusion}
\label{sec:conclusion}

In this paper, we have presented a method based on NEGF to calculate
the equilibrium and non-equilibrium electronic structure of \elvib
interacting single-molecule junctions.  We have applied the method to
a model system which consists of a single electronic level coupled to a
single vibration mode in the central region, the latter in contact
with two non-equilibrium electron reservoirs.

In comparison to previous studies performed within a similar approach
\cite{Mii:2003,Frederiksen:2004b, Galperin:2004b, Mitra:2004,
  Pecchia:2004b, Chen_Z:2005, Ryndyk:2005, Sergueev:2005, Viljas:2005,
  Yamamoto:2005, Cresti:2006, Vega:2006, Egger:2008}, the novelty of
our method lies in the fact that it goes beyond the
conventionally-used self-consistent Born approximation (SCBA).
Higher-order diagrams for the \elvib interacting have been implemented
in our calculations.

In this paper we have considered the second-order diagrams which contain
two-vibron processes: the so-called double exchange DX diagram which is part
of the vertex corrections to SCBA, and the partially dressed DPH diagram
in which the vibron propagator is renormalized by one electron-hole bubble.

We have studied the effects of the first and second-order diagrams on the spectral
functions for a large set of parameters and for different transport
regimes (resonant and off-resonant cases) at equilibrium and in the presence
of a finite applied bias driving the system out of equilibrium.

We have shown the important role played by the Hartree diagram in calculations
based only on first-order diagrams. Such a diagram should not be neglected
unless one works within an applied bias range for which the correspond Hartree
potential is constant.

For calculations including both first and second-order diagram we have
found that the effects of the individual second-order diagram are as
follows: the DX diagram reduces the width of the peaks in the spectral
functions, with an increase of their height, while the DPH has the
opposite effect: it increases the width and decreases the height.
Furthermore, the DX diagram moves the vibron side-band peak position
towards the main peak, and hence gives a better peak separation as
should be obtained from exact calculations.  Calculations performed
with both DX and DPH diagram give intermediate results for the
spectral functions, which are not simply an average/superposition of
the individual effects because of the strong non-linearity involved in
solving the problem self-consistently.

Furthermore, the effects of the second-order diagrams also depend on
the transport regime: at and near equilibrium, the DX diagram
dominates over the DPH diagram in the off-resonant transport regime
(essentially because there are not many electron-hole excitations
available in this regime). In the resonant case, however, both DX and
DPH play an equally important role.  For large non-equilibrium
conditions ($V > \omega$), both second-order diagrams play an
important role, since the corresponding spectral functions look
qualitatively similar to those obtained the resonant regime.

We have thus shown that it is indeed necessary to go beyond SCBA to
obtain the correct results for a wide range of parameters. This has
also been confirmed by comparing our NEGF results to an exact
calculation in terms of \elvib interacting (the multi-channel
inelastic scattering technique MCIST \cite{Ness:2006}), though it
should be noted that comparison between NEGF and MCIST calculations
are only valid for the off-resonant transport regime.

We have also studied in detail the effects of self-consistency on the
calculations, especially for the cases including the second-order
diagrams (see Section~\ref{sec:2ndorder_off_res_equi} and
Figure~\ref{fig:specdens_eq_ofres__strg_HFDXDPH}). We have found a
solution to an old problem, well known in many-body perturbation
theory: in order to avoid negative spectral densities when including
higher order diagrams in the calculations, it seems more appropriate
to start the self-consistent loop with SCBA (Hartree-Fock like)
electron \GFs.

Finally, we have studied the linear response (linear conductance
$G_{\rm lin}$) of the nanojunctions, and have found that in the
off-resonant regime, the value of $G_{\rm lin}$ is governed by the
behaviour of the tail of the main peak at the Fermi level, hence it
depends on both the position and the width of this peak. In the
off-resonant regime, the DPH diagram contribution is negligible and
the DX diagram gives a better polaron shift (better position of the
main peak) and hence the second-order (mostly DX) calculation gives a
better agreement for $G_{\rm lin}$ with the exact result.  For the
near-resonant regime, the contribution of DPH will become more
important and both DX and DPH will play an important role in
determining the value of $G_{\rm lin}$ by changing both the position
and the width of the main peak.

For the resonant regime with electron-hole symmetry, it is necessary
to perform the calculations by including both DX and DPH diagrams in
order to conserve the Fermi-liquid properties of the system.  We
anticipate that this will be also true for higher-order diagrams and
calculations need to be performed by including all diagrams of the
same order.  Calculations performed with partial resummation of a
subset of diagrams will break the expected Fermi-liquid properties of
the system and will probably lead to an incorrect value of the linear
conductance.

Finally, we expect that the effects of the second-order diagrams,
shown in this paper for the spectral functions only, are also
important in the full non-equilibrium transport properties of the
nanojunctions.  The study of these effects on the non-linear
conductance are currently undertaken and will be considered in a
forthcoming paper.

\appendix
\section{Self-energies for the Born approximation and the second-order diagrams}
\label{app:higherorderdiagrams}

In this appendix, we show how to derive the expression for the
self-energies corresponding to one- and two-vibron process
diagrams in terms of \elvib coupling.

Starting from the SSSM Hamiltonian, the general definition of
the one-particle GF of the central region is obtained from 
the time-loop contour-ordered product of the creation and 
annihilation operators in the central region
 \begin{equation}\label{eq:def_GF}
    G(\tau,\tau')= -{\rm i} \expect{T_{C_K} d_H(\tau) d^\dag_H(\tau')}.
 \end{equation}
The time-loop contour $C_K$ contains two branches, the upper $(+)$ and 
the lower $(-)$ branch. 
On the upper branch, time starts in the infinitely remote past
and evolves forwards, then at the turning point, which can be
placed at any arbitrary time, one passes onto the lower branch
where the system evolves backwards in time back to the initially
non-interacting starting point at $t=-\infty$.

Then any expectation value of products of operator reduces to
$\langle\phi_0\vert T_{C_K}\left(\hat{A}(\tau)\hat{B}(\tau')\dots
  S_{C_K} \right)\vert\phi_0\rangle$ where
$\langle\phi_0\vert\dots\vert\phi_0\rangle$ is the average over the
non-interacting ground state.  The operators are then given in the
interaction picture, and $S_{C_K}$ is the generalization of the time
evolution operator on the Keldysh contour
$S_{C_K}=T_{C_K}\left(\exp\{-{\rm i} \int_{C_K}{\rm d}\tau \
  \hat{V}(\tau)\}\right)$ where $T_{C_K}$ is the time-ordering
operator on the contour $C_K$ and $\int_{C_K} {\rm d}t$ implies
integration over $C_K$.  And $\hat{V}$ is the ``perturbation'' to the
reference Hamiltonian, which in our case would be the interaction
\elvib as well as the coupling of the central region to the leads).

Expanding $S_{C_K}$ as a series in terms of the \elvib coupling
Hamiltonian $H_\evib$, one can derive the \elvib self-energies to any
order of the \elvib coupling by calculating any time ordered products
in the series using the usual rules of many-body perturbation theory,
such as Feynmann diagrammatic expansion or Wick's
theorem\cite{Mahan:1990,Abrikosov:1963,Fetter:1971}.

\subsection{Lowest order self-energies}
\label{app:1st_order_diagrams}

The one-vibron process self-energies, corresponding to the Hartree
and Fock-like diagrams (see Figure \ref{fig:FockHartree}) are given by

\begin{equation}\label{eq:SE_Hartree}
\Sigma^H_\evib(\tau_1,\tau_2)=-{\rm i}\gamma_0^2\
D_0(\tau_1,\tau_2)\ G(\tau_1,\tau_1^+) \ ,
\end{equation}
and
\begin{equation}\label{eq:SE_Fock}
\Sigma^F_\evib(\tau_1,\tau_2)={\rm i}\gamma_0^2\
D_0(\tau_1,\tau_2)\ G(\tau_1,\tau_2) \ ,
\end{equation}
where $\tau_i$ are times on the time-loop contour.

The projections onto the real (physical) times are given by,
for example,
\begin{equation}
\Sigma^{F,\zeta_1\zeta_2}_\evib(t_1,t_2)={\rm i}\gamma_0^2\
D_0^{\zeta_1\zeta_2}(t_1,t_2)\ G^{\zeta_1\zeta_2}(t_1,t_2) \ ,
\end{equation}
for the Fock self-energy .

The index $\zeta_i=\pm$ labels the branch of the time-loop 
contour corresponding to the forward ($\zeta_i=+$)/
backward ($\zeta_i=-$) time evolution respectively.

There are several useful relationships between the different projections
(or Keldysh components) of the Green's functions $X \equiv G, D$:
\cite{Keldysh:1965,Craig:1968,MeirWingreen:1992,vanLeeuwen:2006,Rammer:2007}.
They are
\begin{equation}\label{eq:Xr}
X^r=X^{++}-X^{+-}=X^{-+}-X^{--} 
\end{equation}
or equivalently in terms of time-ordered ($t=++$), anti time-ordered ($\tilde t=--$),
greater ($>=-+$) and lesser ($<=+-$) Green's functions:
\begin{equation}\label{eq:Xr_bis}
X^r=X^t-X^<=X^>-X^{\tilde t} 
\end{equation}
and 
\begin{equation}\label{eq:Xa}
X^a=X^{++}-X^{-+}=X^{+-}-X^{--}
\end{equation}
(or equivalently $X^a=X^t-X^>=X^<-X^{\tilde t}$), 
and hence $X^r-X^a=X^>-X^<$.

Similar relationships exist for the self-energies $\Sigma^x \leftrightarrow X^x$
($x=r,a,<,>,t,\tilde t$).

Using these relationships and taking the steady state limit, i.e. $X(t,t')=X(t-t')$,
and after Fourier transformation into an energy representation 
$X(\omega)$, 
we obtain the usual expressions 
\cite{Mii:2003,
Frederiksen:2004b,
Galperin:2004,
Mitra:2004,
Pecchia:2004,
Chen_Z:2005,
Ryndyk:2005,
Sergueev:2005,
Viljas:2005,
Yamamoto:2005,
Cresti:2006,
Vega:2006,
Egger:2008}
for the Hartree $\Sigma^{H,x}_\evib(\omega)$ and Fock 
$\Sigma^{F,x}_\evib(\omega)$
\elvib self-energies ($x=r,a,<,>$).

For example, the Hartree and Fock self-energies are
\begin{gather}\label{eq:Hartree_and_Fock_SE}
\Sigma^{H,x}_\evib(\omega) = -{\rm i}\gamma_0^2\  D_0^x(\omega=0)\ 
\int \frac{{\rm d}\omega'}{2\pi} \ G^<(\omega') \ , \\
\Sigma_{\evib}^{F,y}(\omega)  =  i \gamma_0^2 
\int \frac{d\omega^\prime}{2\pi} \
    D_0^y(\omega^\prime) \
    G^y(\omega - \omega^\prime) \ ,
\end{gather}
where $y$ represents one of the tree Keldysh compoentns $y=t,<,>$. 
Using the relationship $\Sigma^{r}=\Sigma^{t}-\Sigma^{<}$,
we find
  \begin{multline} \label{eq:Fock_r_SE}
      \Sigma_\evib^{F,r}(\omega)  = {\rm i}\gamma_0^2 
      \int \frac{d\omega^\prime}{2\pi} D_0^r(\omega -\omega^\prime)
      G^<(\omega^\prime) \hfill \\
       \qquad+ D_0^<(\omega -\omega^\prime)
    G^r(\omega^\prime) + D_0^r(\omega -\omega^\prime) G^r(\omega) \ . \\
\end{multline}

With the usual definitions for the bare vibron Green's functions
$D_0$:
\begin{gather}\label{eq:D0}
  \begin{split}
    D_0^<(\omega) & = -2\pi i \left[ \expect{N_\ph} \delta(\omega
      -\omega_0)\right. \\
    & \qquad \left. + (\expect{N_\ph} + 1) \delta(\omega+\omega_0) \right],
  \end{split}
  \\
  \begin{split}
    D_0^>(\omega) & = -2\pi i \left[ \expect{N_\ph} \delta(\omega
      +\omega_0)\right. \\
    & \qquad \left. + (\expect{N_\ph} + 1) \delta(\omega-\omega_0) \right],
  \end{split}
  \\
  \begin{split}
    D_0^r(\omega) & = \frac{1}{\omega - \omega_0 +i\eta} \\
    & \qquad - \frac{1}{\omega
      + \omega_0 +i\eta}, \quad \eta\rightarrow 0^+ 
  \end{split}
  \\
  D_0^a(\omega) = [D_0^r(\omega)]^*,
\end{gather}
where $\expect{N_{\text{ph}}}$ is the averaged number of excitations in
the vibration mode of frequency $\omega_0$ given by the Bose-Einstein 
distribution at temperature $T_{\rm vib}$.  

One can see that $\Sigma^{H,<,>}_\evib(\omega)=0$
because $D_0^{<,>}(\omega=0)=0$ unless $\omega_0=0$ which would be
an odd case of study.
And since $D_0^{r,a}(\omega=0)=-2/\omega_0$, one has
\begin{equation}\label{eq:Hartreeselfenergy}
  \Sigma_\evib^{H,r} =  \Sigma_\evib^{H,a} =  \frac{\gamma_0^2}{\omega_0} 2 {\rm i} \int
  \frac{d\omega^\prime}{2\pi} G^<(\omega^\prime) \ .
\end{equation}

Furthermore the lesser and greater Fock self-energies can be expressed
in a more compact form:
\begin{equation}
  \label{eq:SelfEnergylesser_and_greater}
 \begin{split}
 \Sigma_{\evib}^{F,\{<,>\}}(\omega) & = \gamma_0^2 \left[ \expect{N_\ph}
  G^{\{<,>\}}(\omega \mp \omega_0) \right. \\
  & \quad \left. + (\expect{N_\ph} + 1) G^{\{<,>\}}(\omega \pm \omega_0) \right].
\end{split}
\end{equation}

\subsection{Second-order self-energies}
\label{app:2nd_order_diagrams}

The two-vibron process self-energies correspond to the two diagrams
shown in Figure \ref{fig:DX-DPH}, i.e. the so-called double
exchange (DX) diagram and GW-like (dressed phonon/vibron DPH) diagram in which 
the vibron propagator is renormalized by a single electron-hole bubble
polarization.

The expressions for the self-energies for these diagrams are:
\begin{equation}
  \label{eq:Dex_SE}
  \begin{split}
    \Sigma^{DX}_\evib(\tau_1,\tau_2)  = - \frac{\gamma_0^4}{3}
    \int_{C_K} &  {\rm d}\tau_3 {\rm d}\tau_4 
    G(\tau_1,\tau_3) D_0(\tau_1,\tau_4) \times \\ 
    & G(\tau_3,\tau_4) D_0(\tau_3,\tau_2) G(\tau_4,\tau_2), 
  \end{split}
\end{equation}
and
\begin{equation}\label{eq:GWlike_SE}
  \begin{split}
    \Sigma^{DPH}_\evib(\tau_1,\tau_2)  = 
    + \frac{\gamma_0^4}{3} G(\tau_1,\tau_2) \times \\
    \int_{C_K}  {\rm d}\tau_3 {\rm d}\tau_4 
    D_0(\tau_1,\tau_3) G(\tau_3,\tau_4) G(\tau_4,\tau_3) D_0(\tau_3,\tau_2). 
  \end{split}
\end{equation}
The factor $1/3$ comes from the series expansion of the exponential
in the time evolution operator $S_{C_K}$ and the fact that one obtains
8 equivalent diagrams of the $\gamma_0^4$ order.

Taking the steady state limit and after Fourier transformation,
the different Keldysh components of the double-exchange (DX) self-energy 
are given by
\begin{equation}\label{eq:Dex_SE-bis}
\begin{split}
\Sigma^{DX,\zeta_1\zeta_2}_\evib(\omega)  = -\frac{\gamma_0^4}{3}
\int \frac{{\rm d}u}{2\pi} 
\frac{{\rm d}v}{2\pi} 
\sum_{\zeta_3,\zeta_4} \zeta_3 \zeta_4 G^{\zeta_1 \zeta_3}(v) \times \\
D_0^{\zeta_i \zeta_4}(\omega-v)
G^{\zeta_3 \zeta_4}(v-u) D_0^{\zeta_3 \zeta_2}(u) G^{\zeta_4 \zeta_2}(\omega-u).
\end{split}
\end{equation}

\subsection{Normalisation of the vibron propagators}
\label{app:dressing_D0}

The expression for the self-energy $\Sigma^{DPH}_\evib(\tau_1,\tau_2)$
can actually be recast in a Fock-like diagram with a renormalised
vibron propagator $\mathcal{D}_0$ \cite{Galperin:2004,Viljas:2005}:
\begin{equation}\label{eq:dressedPH_Fock}
\Sigma^F_\evib(\tau_1,\tau_2)={\rm i}\gamma_0^2\
\mathcal{D}_0(\tau_1,\tau_2)\ G(\tau_1,\tau_2) \ ,
\end{equation}
where
the dressed vibron $\mathcal{D}_0(\tau,\tau')$ is given by
\begin{equation}\label{eq:dressedPH}
\mathcal{D}_0(\tau,\tau') = \gamma_0^2\ \int_{C_K}  {\rm d}\tau_2 {\rm d}\tau_3 \
    D_0(\tau,\tau_3)\ \Pi(\tau_3,\tau_2)\ D_0(\tau_2,\tau') 
\end{equation}
and
the polarization $\Pi(\tau,\tau')$ is given by the electron-hole bubble diagram:
\begin{equation}\label{eq:polarisation}
\Pi(\tau,\tau') = -{\rm i}\ G(\tau,\tau') \ G(\tau',\tau) \ .
\end{equation}

So, in principle, we already have all the ingredients to perform calculations
using the fully dressed vibron propagator $\mathcal{D}(\omega)$ \cite{Galperin:2004,Viljas:2005}
whose advanced and retarded components are given by 
\begin{equation}\label{eq:fullydressedPH}
\mathcal{D}^{\{r,a\}}(\omega)=
\left[ 
[D_0^{\{r,a\}}(\omega)]^{-1} - \gamma_0^2\ \Pi^{\{r,a\}}(\omega) 
\right]^{-1} \ .
\end{equation}
 
The vibron spectrum is renormalized by the polarization. Ther is a shift of the vibron
energy related to $\gamma_0^2\Re e \Pi^{\{r,a\}}$ and a finite linewidth of the peaks 
related to $\gamma_0^2\Im m \Pi^{\{r,a\}}$, instead of Dirac $\delta$ peaks 
as in Eq.(\ref{eq:D0_PVanddelta}).

We have not performed the full vibron renormalization in the present work, because we 
want to study the effects of different diagrams of the same order, and order by order.
$\mathcal{D}(\omega)$ correspond to a partial resummation of all the
bubble diagrams and contains all (even) orders of the \elvib coupling.
The effect of this  GW-like \elvib interaction will be
presented in a forthcoming paper.

\section{Details of numerical calculations}
\label{app:numericalbits}

In this section we provide more details about how calculations for the self-energies
are performed.

\begin{figure}
  \centering
  \includegraphics[width=8cm]{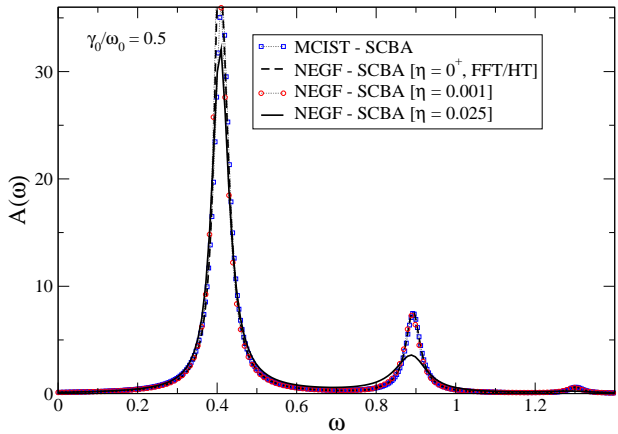}
  \caption{Equilibrium spectral functions for the off-resonant regime
    calculated with MCIST and NEGF-SCBA.  
    MCIST calculations are performed at the Hartree-Fock level (BA).
    NEGF are done using the limit $\eta\rightarrow 0^+$ for 
    the imaginary part of the bare vibron GF and also by using a tiny, 
    but finite value for $\eta$ (0.001 and 0.025). See main text for
    details.  
    The other parameters are $\varepsilon_0=+0.5$ (electron
    transport), $\gamma_0=0.20$ (upper panel), $\omega_0=0.40,
    t_{0L}=t_{0R}=0.15$.
    For a small enough value of $\eta$, one recovers the exact MCIST
    as well as the $\eta\rightarrow 0^+$ NEGF results obtained from
    Hilbert transform (HT). Larger values of 
    $\eta$ widen the peaks; however the spectroscopic information
    remains. For such values of $\eta$, one can work with fewer grid
    points and makes second-order self-energy calculations tractable.}
    \label{fig:app:specdens_MCISTvsSCBAvsfullSC}
\end{figure}

First, we start by using the following decomposition of the retarded and advanced
vibron propagator $D_0^{r,a}$:
\begin{equation}
  \label{eq:D0_PVanddelta}
  \begin{split}
    D_0^{r,a}(\omega) & = \frac{1}{\omega - \omega_0 \pm {\rm i}\eta} - \frac{1}{\omega
      + \omega_0 \pm {\rm i}\eta}, \quad \eta\rightarrow 0^+ \\
    & = \mathrm{P.V.} \frac{1}{\omega - \omega_0} - \mathrm{P.V.}
    \frac{1}{\omega + \omega_0} \\
    & \qquad \mp i\pi(\delta(\omega -\omega_0) -
    \delta(\omega + \omega_0),
  \end{split}
\end{equation}
and the definition of the Hilbert transformation (H.T.):
\begin{equation}
  \label{eq:HilbertTransformDefinition}
  \mathcal{H}(y) = \mathrm{H.T.}[f(x)] = \frac{1}{\pi} \mathrm{P.V.}
  \int_{-\infty}^\infty dx \frac{f(x)}{x - y}.
\end{equation}

One can then rewrite the Fock retarded (advanced) self-energy in terms 
of retarded (advanced) and lesser electron Green's functions only:
\begin{widetext}
\begin{equation}
  \label{eq:electronvibron-fock-selfenergy-copy}
 \begin{split}
   \Sigma_\evib^{F,r}(\omega) = & \gamma_0^2  \biggl[ \expect{N_\ph}
    G^r(\omega -\omega_0) + (\expect{N_\ph} + 1) G^r(\omega +\omega_0)
     + \frac{1}{2}\left( G^<(\omega -\omega_0) + G^r(\omega
      -\omega_0) - G^<(\omega + \omega_0) - G^r(\omega + \omega_0)
    \right) \\
    & \quad + \frac{i}{2} \left( \mathcal{H}[G^< + G^r](\omega
      +\omega_0) - \mathcal{H}[G^< + G^r](\omega - \omega_0) \right) \biggr],
  \end{split}
\end{equation}
and
\begin{equation}
  \label{eq:electronvibron-fock-selfenergy-advanced}
 \begin{split}
   \Sigma_\evib^{F,a}(\omega) = & \gamma_0^2  \biggl[ \expect{N_\ph}
    G^a(\omega -\omega_0) + (\expect{N_\ph} + 1) G^a(\omega +\omega_0)
     - \frac{1}{2}\left( G^<(\omega -\omega_0) - G^a(\omega
      -\omega_0) - G^<(\omega + \omega_0) + G^a(\omega + \omega_0)
    \right) \\
    & \quad + \frac{i}{2} \left( \mathcal{H}[G^< - G^a](\omega
      +\omega_0) - \mathcal{H}[G^< - G^a](\omega - \omega_0) \right) \biggr].
  \end{split}
\end{equation}

\end{widetext}

These expressions are in a form convenient for computation,
since, as for $\Sigma_\evib^{F,<,>}$, they involve the direct evaluation
of $G^{r,a,<}$ on a energy grid shifted by $\pm\omega_0$ in addition
with the Hilbert transform of $G^<$ and $G^{r,a}$ on a shifted energy grid.

The Hilbert transformation is actually a conventional convolution
product of the trial function with an appropriate $(1/x)$-like kernel.
Calculation of convolution products can be made faster, instead of
scaling as the square of the number of grid points, by the use of FFT
routines.  However since FFT routines introduce artifical periodic
boundary conditions, and one has to make sure that the functions, which 
decay slowly as $1/\omega$ (the kernel of the Hilbert transform,
and real part of $G^{r,a}$), have sufficiently small values at the
boundaries of the energy grid. Otherwise, the corresponding
discontinuities introduce spurious oscillations in the FFT.  This means
that one has to work with a wide energy grid and with a correspondingly
large number of grid points to keep a good enough resolution.

Working with a large number of grid points $N_{\rm pts}>10^4$ is not
really a problem when calculating convolution products with FFT, since
the number of operations reduces from $N_{\rm pts}^2$ to $N_{\rm
  pts}\log_2 N_{\rm pts}$, or even when without using FFT routines.
However, it is not possible to reduce the calculation of the
second-order DX self-energy to some sort of convolution product, and
the number of operations then scales as $N_{\rm pts}^3$. Such a
scaling starts to make calculations seriously impractical when
working with a large number of grid points.

Hence, we have adopted another stategy which consists of introducing, in
the definition of $D_0^{r,a}$, a tiny but finite imaginary part $\rm i\eta$
instead of taking the limit $\eta\rightarrow 0^+$. This is just a natural
step towards working with a fully dressed (renormalized) vibron propagator
as already discussed in Section \ref{app:dressing_D0}.

By using this tiny but finite imaginary part $\rm i\eta$, we avoid having
to deal numerically with the Dirac $\delta$-function and consequently with the Hilbert
transform and all the problems associated with the slowly decaying kernel.
We can then work with a less wide energy grid and fewer points
while keeping a good energy resolution. Hence the calculations of the second
order DX self-energy become more tractable.
Of course, there is a price to pay for that: there is a lowest bound for the
possible values of $\eta$, and this lowest bound is strongly linked to the value
of the grid spacing. By trial and error, we have found that $\eta$ must be 
at least equal to 2 or 3 times the grid spacing.
 
Now we have to check and compare the results for the spectral
functions obtained from these two different methods of calculation.
Figure \ref{fig:app:specdens_MCISTvsSCBAvsfullSC} show the spectral
functions for the off-resonant regime at equilibrium and for
weak/intermediate \elvib coupling.  Calculations have been performed
with MCIST at the Hartree-Fock level (MCIST-SCBA) which serves as a
reference calculation. The NEGF-SCBA calculations performed within the
limit $\eta\rightarrow 0^+$ by using Hilbert transform and FFT for the
Fock self-energy (in the present case with a 131072 grid points
ranging from -30 to +30)  are identical to the MCIST-SCBA results as
expected.  The NEGF-SCBA calculations performed with a tiny but finite
imaginary part $\rm i\eta$ have been done for different sets of
parameters. Only two are shown in Figure
\ref{fig:app:specdens_MCISTvsSCBAvsfullSC}: $\eta=0.001$ corresponding
to a energy grid ranging from -10 to +10 and with 32769 points, and
$\eta=0.025$ corresponding to a energy grid ranging from -10 to +10
and with 2049 points.  The spectral function obtained with
$\eta=0.001$ is virtually identical to the corresponding ``exact''
calculations, while the spectral function obtained with $\eta=0.025$
has slightly broadened peaks with a reduced amplitude, even though such a
calculation has been performed with only 2049 grid points---which is in
the range of good values of grid points to make calculation of the second-order DX
self-energy numerically tractable.  In any case, the most important point is
that the spectral information (peak positions) are not dependent on
the finite value of $\eta$.

Now we also have to verify the influence of $\eta$ on another transport regime
for which we do not have any exact reference calculations, but for which we
may expect to recover conventional properties of an interacting Fermi liquid
at very low temperatures.

\begin{figure}
  \includegraphics[width=8cm]{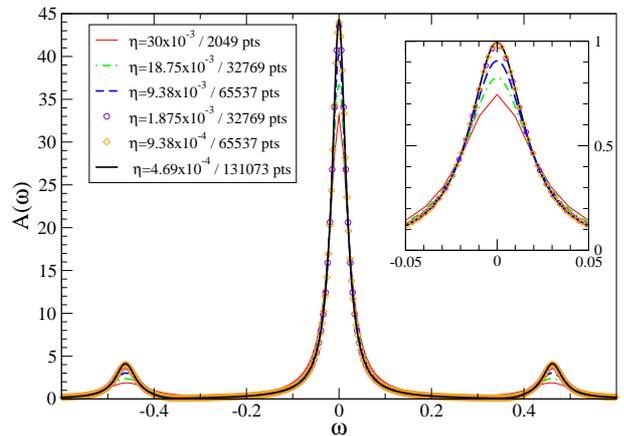}
  \caption{(Color online). Equilibrium spectral functions for the resonant regime
    obtained within NEGF (Fock self-energy only).
    Calculations are performed for different values of the imaginary part $\eta$ 
    (0.03, 0.01875, 0.00938, 0.001875, 0.000938) and for different number of 
    energy grid points (2049, 32769, 65537, 131073).
    The energy grid is ranging from -10 to +10.
    The inset shows a zoom around $\omega=0$ of the rescaled spectral functions, 
    giving then the linear conductance $G_{\rm lin}$ at $\omega=0$.
    The other parameters are $\varepsilon_0=0.0$, $\gamma_0=0.23$ , 
    $\omega_0=0.40, t_{0L}=t_{0R}=0.15$.
    Peaks in the spectral functions narrow with decreasing values of $\eta$ as
    expected. And one recovers a perfect 
    linear conductance $G_{\rm lin}=G_0$ at $T=0$.}
    \label{fig:app:specdens_NEGFresSCBA_vs_etaD0+Nbpts}
\end{figure}

Figure \ref{fig:app:specdens_NEGFresSCBA_vs_etaD0+Nbpts} shows the equilibrium 
spectral functions for the resonant regime. Calculations are performed with
only the Fock self-energy and for different values of $\eta$ 
(0.03, 0.01875, 0.00938, 0.001875, 0.000938) and for different number of 
energy grid points (2049, 32769, 65537, 131073).
The energy grid is ranges from -10 to +10.
As expected, the main central peak and the vibron side-band peak narrow with
increasing amplitude when the value of $\eta$ decreases, until reaching an
asymptotic behaviour for $\eta \sim 10^{-3}$.

The most interesting result is given by the behaviour of spectral
function at the equilibrium Fermi level $\mu^{\rm eq}=0$.  Such a
behaviour is exemplified in the inset of Figure
\ref{fig:app:specdens_NEGFresSCBA_vs_etaD0+Nbpts}.  The inset shows
the spectral function rescaled by $2\Gamma(\mu^{\rm eq})$ so that the
value at $\omega=0$ corresponds to the linear conductance $G_{\rm
  lin}$ (see Section \ref{sec:linear_reponse}).  One can see that
$G_{\rm lin}$ goes towards the unit of conductance as $\eta
\rightarrow 0$. Such a limit corresponds to a perfect conductance, as
obtained for the corresponding non-interacting system.

This result shows that even in the presence of \elvib interactions, the linear
conductance is perfect, as one also obtains for the non-interacting case.
The \elvib interaction (as least at the Fock level) does not renormalize 
the linear conductance.
For the fully electron-hole symmetric system (no Hartree self-energy), we have 
checked that the value of $\Im m \Sigma^r_\evib(\omega)$ is zero at $\omega=0$ 
and at $T=0$ (at equilibrium). The real part of the interacting self-energy
$\Re e \Sigma^r_\evib=0$ at $\omega=0$ by definition because it is an odd 
function of $\omega$.
Our \elvib interacting system conserves its Fermi-liquid properties at zero
temperature and equilibrium, as also obtained for interacting electron systems
(see for example Refs.[\onlinecite{Ferretti:2005a,Ferretti:2005b,Cornaglia:2004,Oguri:1997}]).
We expect that such a behaviour also hold for other \elvib diagrams than the
Fock diagram.
Out of equilibrium, the system may lose its conventional Fermi-liquid properties,
as we have already shown in the main part of the paper.

In conclusion, we want to stress that by introducing a finite, but tiny, value for
the imaginary part $\eta$, one might not get the exact linear conductance value.
However this is a numerical artifact which can be controlled by increasing 
the number of energy grip points and reducing the value of $\eta$.
Nonetheless we expect that the relative effects of different \elvib diagram are correctly
obtained from our calculations, whatever the small values of $\eta$ are.
And of course, in order to perform a correct analysis of such effects, one should
only compare results for either spectral functions or linear conductance obtained numerically 
with the same value of $\eta$.

\begin{acknowledgments}

  We would like to thank Martin Stankovski for useful discussions of the
  second-order diagrams and Ulf von Barth for discussion of the
  negative DOS problem.  This work was funded in part by the European
  Community's Seventh Framework Programme (FP7/2007-2013) under grant
  agreement no 211956 (ETSF e-I3 grant).  RWG was also supported in
  part by the National Science Foundation under Grant No.PHY05-51164.

\end{acknowledgments}

\end{document}